\documentclass[11pt]{article}
\usepackage{xcolor}
\usepackage{graphicx}
\usepackage{float}
\usepackage{amsthm,comment}
\usepackage{url}
\usepackage[margin=1in]{geometry}
\usepackage[ruled,vlined]{algorithm2e}
\usepackage{enumitem}
\usepackage{natbib}
\RequirePackage{amsthm,amsmath,amsfonts,amssymb}
\usepackage[ruled,vlined]{algorithm2e}
\usepackage{setspace}
\usepackage{tabularx} 
\usepackage{booktabs}
\usepackage{authblk}

\newtheorem{theorem}{Theorem}[section]
\newtheorem{lemma}[theorem]{Lemma}
\theoremstyle{remark}

\usepackage[colorlinks,citecolor=blue,urlcolor=blue]{hyperref}
\usepackage{caption}
\usepackage{subcaption}  
\usepackage{multirow}
\begin{document}

\title{Identifying latent groups in spatial panel data using a Markov random field constrained product partition model}
\author[1]{Tianyu Pan}
\author[2]{Guanyu Hu}
\author[1]{Weining Shen}
\affil[1]{Department of Statistics, University of California, Irvine}
\affil[2]{University of Missouri - Columbia, Columbia, MO, 65211}

\maketitle

\begin{abstract}
Understanding the heterogeneity over spatial locations is an important problem that has been widely studied in many applications such as economics and environmental science. In this paper, we focus on regression models for spatial panel data analysis, where repeated measurements are collected over time at various spatial locations. We propose a novel class of nonparametric priors that combines Markov random field (MRF) with the product partition model (PPM), and show that the resulting prior, called by MRF-PPM, is capable of identifying the latent group structure among the spatial locations while efficiently utilizing the spatial dependence information. We derive a closed-form conditional distribution for the proposed prior and introduce a new way to compute the marginal likelihood that renders efficient Bayesian inference. We further study the theoretical properties of the proposed MRF-PPM prior and show a clustering consistency result for the posterior distribution.  We demonstrate the excellent empirical performance of our method via extensive simulation studies and applications to a US precipitation data and a California median household income data study.

\end{abstract}

\newpage
\section{Introduction}\label{sec 1: Introduction}

Panel data has been widely studied in many applications such as economy \citep{pesaran2015time} and climate science \citep{hao2016influence} since it represents a common data format where observations are collected for each study subject at different time points. Our focus in this paper is to model a special type of panel data, called by {\it spatial panel data}, where each study subject represents a spatial location and there is a need to take account for the spatial dependence among those locations. Spatial panel data analysis has received a growing interest in recent years \citep{pesaran2011large,elhorst2014spatial,belotti2017spatial} and a central question is to model the relationship between variables measured repeatedly over a study time period at various spatial locations. For example, in economic studies, it is of interest to quantify the association between the median household income and other economic indicators such as gross domestic product (GDP) and unemployment rate over time. In environmental studies, understanding the effect of greenhouse gas emissions on climate change is an important research direction. 

The aforementioned question can be naturally formulated as a regression problem in statistics; and it is now well recognized that the regression parameters (e.g., coefficients and variance) can be {\it highly variable} across different spatial locations \citep{hsiao1997panel,browning2007heterogeneity, su2013testing}. To account for such spatial heterogeneity pattern, it is common to assume a {\it latent group structure}, i.e., spatial locations are grouped into clusters and those assigned to the same cluster share the same set of regression parameters. This strategy has several advantages in practice. First, for the obvious reason that neglecting the unobserved heterogeneity may lead to inconsistent parameter estimation and severally misleading results as demonstrated by the famous Simpson's paradox and other examples for spatial panel data \citep{wagner1982simpson, su2016identifying,hsiao2014analysis}. Secondly, the obtained latent group structure is usually informative for empirical analysis, such as finding possible unobserved confounders and performing secondary analysis. Another benefit is that the latent group structure for spatial panel data in general tends to coincide with the geographic locations of the study  \citep{miao2020panel}, which automatically provides a useful way to incorporate the spatial dependence information and hence improve the efficiency/accuracy of the studied model.  

Several approaches have been introduced in the frequentist literature to study panel data regression model with latent group structure. For example, \citet{lin2012estimation}  considered a panel data linear regression model with group-varying slopes. This method was further extended to allow both group-varying intercept and slopes by \citet{su2016identifying}. Several more complicated models have also been proposed with different features, e.g., group-specific time patterns \citep{bonhomme2015grouped}, time-varying grouped coefficients \citep{su2019sieve} and group-varying threshold variables \citep{miao2020panel}. Despite the success of those frequentist approaches, limited effort has been made under the Bayesian framework until very recently \citep{zhang2020forecasting}. Conceptually, an ideal Bayesian approach would naturally be able to incorporate the spatial dependence and latent group structure information in the prior distribution. The inference may also be conveniently conducted without pursing complicated procedures such as bootstrap or post model selection. The main goal of our paper is to pursue under this direction by introducing a new class of nonparametric priors and exploring their computational and theoretical properties.


Our first step towards constructing a prior distribution for spatial panel data with group structure is to recognize that 
a latent group structure is essentially equivalent to a \emph{partition} of spatial locations. Therefore we only need a class of priors assigned to the space of partitions, and this is usually achieved by specifying a class of partition probability functions. Among this class, the product partition model (PPM), which was first introduced by \citet{hartigan1990partition} and studied from a Bayesian point of view by \citet{quintana2003bayesian}, has received a considerable interest over the years. PPM is defined by taking the product of some non-negative cohesion functions $h(c)$ over different clusters, where $h(c)$ measures the similarity between individual subjects assigned to the same cluster $c$ \citep{design1978fundamentals}. It has been shown that the PPM prior has strong connections to the marginal prior on partitions induced by the Dirichlet Process prior (DP, \citet{green2001modelling}) and the Mixture of Finite Mixture Model prior (MFM, \citet{miller2018mixture}). 
Lately, PPM was extended to include covariate  \citep{park2010bayesian, page2015predictions} and spatial information \citep{page2016spatial}. But it still remains unclear how to {\it formally} incorporate the spatial dependence information into the PPM.

To solve this issue, in this paper, we introduce a new class of priors called by a Markov random field constrained product partition model (MRF-PPM) prior. This prior is generated by taking the product of two priors, a Markov Random Field (MRF) prior and a PPM prior. There is a long history of using MRF priors  defined on undirected graphs in the literature to capture the local homogeneity in image segmentation, spatial statistics and Bayesian nonparamerics literature  \citep{geman1984stochastic,orbanz2008nonparametric,blake2011markov}. However, to the best of our knowledge, MRF-PPM prior, as a general class of priors that combines MRF and PPM, has not been systemically studied in the literature in terms of its theoretical and computational properties; and it is our goal to fill this gap in this paper. 


We discuss the outline for the rest of this paper and highlight our main contributions as follows. First, we give a formal definition of the MRF-PPM prior and study its fundamental properties to bridge the connection between this prior and those in the literature by showing that several commonly used nonparametric priors \citep{zhao2020bayesian,hu2020bayesian,orbanz2008nonparametric} are special cases of MRF-PPM. Next, we derive a closed-form full conditional distribution formula for MRF-PPM prior that renders tractable posterior computation. We also introduce a Monte Carlo estimation method for computing the marginal likelihood that avoids the Pseudo-Bias issue as  pointed out in \citet{lenk2009simulation}. Thirdly, we provide a theoretical guarantee for MRF-PPM by showing a clustering consistency result stating that with posterior probability tending to one, the posterior distribution of MRF-PPM is capable of identifying the correct unknown partition structure in the spatial panel data. This result is applicable to any regression models with well-defined posterior contraction rate under mild identifiability conditions. Finally, we demonstrate the excellent numerical performance of the proposed MRF-PPM method via simulation studies and two real data application examples.

\section{Methodology}\label{sec 2: Methodology}

\subsection{Markov Random Field Constrained PPM}\label{sec 2.1: MRF-PPM}
Consider a total of $N$ spatial locations. For location $i$, suppose that we observe a response $Y_{i}(t^{(i)}_j)$ and a $p$-dimensional covariate vector $X_{i}(t^{(i)}_j)$ at time point $t^{(i)}_j$, for $j=1,\ldots,n_i$, where $n_i$ is the total number of time points observed for location $i$. We use $c_i$ to denote the cluster assignment for location $i$, and for those locations that belong to the same cluster index set $c$, i.e., $i \in c$, we use $\theta_c$ to denote the common set of modeling parameters being shared within the cluster $c$. Therefore, our spatial panel data regression model with latent groups structure can be written in the following way, 
\begin{equation}\label{eq2}
    \begin{split}
     & Y_{i}(t^{(i)}_j)\mid X_{i}(t^{(i)}_j) \sim f_{\theta_c}(Y_{i}(t^{(i)}_j)\mid X_{i}(t^{(i)}_j))\text{, for $j=1,\ldots,n_i$,~\text{and} $i \in c$},
    \end{split}
\end{equation}
where $f_{\theta_c}$ is the regression likelihood function for cluster $c$. For the rest of the paper, we also use  $Y_i(t)$ to denote the observation collected at time $t$ for location $i$ for the simplicity of the notation. Note that model \eqref{eq2} allows the temporal correlation between $Y_i(t^{(i)}_j)$ and $Y_i(t^{(i)}_k)$ for every $k\neq j$. To model the clustering structure, we consider a prior on the partition of the index set $[N] = \{1,\ldots,N\}$ and the associated parameters sets $\theta_c$ as follows, 
\begin{align*}
 \theta_c\stackrel{\text{i.i.d}}{\sim} G_0\text{, for $c\in \mathcal{C}$},~~ \mathcal{C}\sim p(\mathcal{C}),
\end{align*}
where $G_0$ is a non-atomic base measure for $\theta_c$ with a probability density function $g(\cdot)$, $\mathcal{C}$ is a partition of $[N]$, and $p(\mathcal{C})$ is a probability mass function over $\mathcal{C}$. It is common to consider a product partition model (PPM) for $p(\mathcal{C})$, that is, 
\begin{equation}\label{eq3}
    \begin{split}
     & p(\mathcal{C})\propto \Pi_{c\in\mathcal{C}}h(c),
    \end{split}
\end{equation}
where $h(c)\geq 0$ is the cohesion function that measures the similarity between individual units assigned to the same cluster $c$.

To account for spatial correlation among different locations. We propose to incorporate a Markov random field (MRF) structure on $p(\mathcal{C})$. Consider a collection of parameters $\{\theta_1,\theta_2,\ldots,\theta_N\}$ defined on an undirected known graph $\mathcal{G}_N=(V_{\mathcal{G}_N},E_{\mathcal{G}_N})$, where $V_{\mathcal{G}_N}=\{\theta_1,\theta_2,\ldots,\theta_N\}$ is the vertex set and $E_{\mathcal{G}_N}$ is the set of edges. Recall that in our case, $\theta_i$ is the regression parameter for location $i$, which is equivalent to $\theta_{c_i}$ defined in \eqref{eq2}. Given the graphical information, a joint distribution $m$ on $V_{\mathcal{G}_n}$ is called an MRF w.r.t $\mathcal{G}_N$ if
\begin{equation}\label{eq4}
    \begin{split}
     & m(\theta_i\mid\theta_{(-i)};\mathcal{G}_N)=m(\theta_i\mid\theta_{\partial{(i)}};\mathcal{G}_N),
    \end{split}
\end{equation}
where $\partial{(i)}=\{j\mid (i,j)\in E_{\mathcal{G}_N}\}$, $\theta_{(-i)}= \{\theta_i\}_{i=1}^N\setminus\{\theta_i\}$, and, $\theta_{\partial{(i)}}= \{\theta_j\mid (i,j)\in E_{\mathcal{G}_N}\}$. This Markov property indicates that $\theta_i$'s distribution only depends on its neighbors, i.e., vertexes that are connected to $\theta_i$. We can further define a MRF joint cost function, which is not necessarily a probability density function, as $M(\theta_1,\ldots,\theta_N\mid\mathcal{G}_N)=\Pi_{c\in\mathcal{C}}l(\theta_c)k(c\mid\mathcal{G}_N)$ (sometimes denoted by $M$) that satisfies
\begin{equation}\label{eq5}
     k(c\cup\{i\}\mid\mathcal{G}_N)=k(c\mid\mathcal{G}_N)\cdot k_i(\partial{(i)}\cap c\mid\mathcal{G}_N)\text{, for every $i$ and every $c \subset [N]$},  
\end{equation}
where $l(\cdot)$ is a non-negative function, $k(\cdot\mid\mathcal{G}_N)$ and $k_i(\cdot\mid\mathcal{G}_N)$ are non-negative cohesion functions defined for every $c\subseteq [N]$ given the graphical information; and it satisfies  $k(\{i\}\mid\mathcal{G}_N)=1$ for all $i$, and $k(\emptyset\mid\mathcal{G}_N)=k_i(\emptyset\mid\mathcal{G}_N)=1$. By letting $P(\theta_1,\ldots,\theta_N)$ ($P$ for short) be the prior on $\{\theta_1,\ldots,\theta_N\}$ defined in \eqref{eq2}, which is proportional to $\Pi_{c\in\mathcal{C}}g(\theta_c) h(c)$, we can construct a MRF-PPM prior $\Pi$ by taking the product of $P$ and the MRF cost function $M$ with some normalizing constant $K_0$ as follows,
\begin{equation}\label{eq6}
    \begin{split}
     & \Pi(\theta_1,\ldots,\theta_N\mid\mathcal{G}_N)=K_0 M(\theta_1,\ldots,\theta_N\mid \mathcal{G}_N)P(\theta_1,\ldots,\theta_N).
    \end{split}
\end{equation}
It can be shown that the proposed MRF-PPM prior enjoys the following three attractive properties:
\begin{enumerate}
\item[(P1)] If $l(\theta) g(\theta)$ is integrable as a function of $\theta$, then $\Pi(\cdot\mid\mathcal{G}_N)$ is still a product partition model, with cohesion function equals to $k(\cdot\mid\mathcal{G}_N)h(\cdot)$ and probability density function of base measure being  $K_1 l(\cdot)g(\cdot)$ for some normalizing constant $K_1$.
\item[(P2)] It inherits the ability of clustering because it provides a full support over the entire space of partitions.
\item[(P3)] It is exchangeable since the cohesion function is 
invariant under permutation (it only depends on the clustering configuration), which by De Finetti's theorem \citep{de1929funzione} justifies the existence of the MRF-PPM prior.
\end{enumerate}
Next we derive the full conditional distribution of MRF-PPM prior in Theorem \ref{thm1}. The proof is given in the Appendix.
\begin{theorem}\label{thm1}
Suppose that $l(\theta)g(\theta)$ in MRF-PPM prior is integrable as a function of $\theta$, then  the conditional distribution of $\theta_i$ given $\theta_{(-i)}$, induced partition $\mathcal{C}_i$, and distinct values $\{\theta_c\}_{c\in\mathcal{C}_i}$, is proportional to
\begin{equation}\label{eq7}
    \begin{split}
     & \frac{k(\{i\}\mid\mathcal{G}_n)h(\{i\})}{K_1}L_0+\sum_{c\in\mathcal{C}_i}k_i(\partial{(i)}\cap c\mid\mathcal{G}_n)\frac{h(c\cup \{i\})}{h(c)}\delta_{\theta_c},~\text{for every}~i,
    \end{split}
\end{equation}
where $L_0$ is the base measure associated with the probability density function $K_1l(\theta)g(\theta)$.
\end{theorem}
Apparently, if $l(\theta)=1$, we have the base measure $L_0=G_0$. From the second term in \eqref{eq7}, we can see that  MRF-PPM is able to account for the spatial correlation since location $i$ would have a higher probability of being assigned to a specific cluster that includes more of its neighbors. That probability is determined by the function $k_i(\partial{(i)}\cap c\mid\mathcal{G}_n)$, which satisfies the Markov property in \eqref{eq5}. 

In addition to PPM, we can also impose an MRF structure on an \emph{exchangeable partition probability function} (EPPF) \citep{pitman2002combinatorial} as described in the following theorem.  
\begin{theorem}\label{thm2}
If the partition probability function of $P$ is an EPPF, and the cluster-wise parameters are i.i.d sampled from a base measure $G_0$, then the resulting MRF-EPPF satisfies Properties (P1)-(P3) and a full conditional distribution can be obtained similarly with \eqref{eq7}.
\end{theorem}
Theorem \ref{thm2} is widely applicable to many commonly used priors in Bayesian nonparametrics literature. For example, it is well known that the partition probability function of Dirichlet process is an EPPF. Also, partition probability function of mixture of finite mixture (MFM) prior is an EPPF  \citep{miller2018mixture}. Therefore the MRF structure can be convininently combined with those two priors. 

It is also worthy mentioning that under the MRF structure, $\Pi(\theta_1,\ldots,\theta_{N-1}\mid\mathcal{G}_{N-1})\neq \int \Pi(\theta_1,\ldots,\theta_{N}\mid\mathcal{G}_{N}) d\theta_N$, because the new observation $\theta_n$ will provide extra spatial information to the historical data $\{\theta_i\}_{i=1}^{N-1}$, hence the marginal distribution of $\{\theta_i\}_{i=1}^{N-1}$ will change. As a consequence, the Kolmogorov's extension theorem \citep{durrett2019probability} cannot be directly applied to show the existence of $\Pi(\theta_1,\ldots,\theta_{N}\mid\mathcal{G}_{N})$ when $N\to\infty$. For the same reason, the Polya urn scheme is not available for MRF-MFM. However, this will not affect our method because we only focus on the fixed $N$ situation, that is, the number of spatial locations of interest is fixed in the study. 

\subsection{Model Specification}\label{sec 2.2: Model Specification}
Next we focus on the linear regression case with Gaussian errors and demonstrate how the proposed prior works for the  model introduced in \eqref{eq2}. The full model can be formulated in the following hierarchical order, 
\begin{equation}\label{eq8}
    \begin{split}
     & Y_{i}(t^{(i)}_j)\mid \left\{ e_{i}(t^{(i)}_j),X_{i}(t^{(i)}_j) \right\}\stackrel{\text{ind}}{\sim} \mathcal{N}(X_{i}(t^{(i)}_j)\beta_c+e_{i}(t^{(i)}_j),\sigma_c^2\cdot \alpha_c),\\
     & e_{i}(t^{(i)}_j)\sim \mathcal{N}(0,K_{\sigma_c,\ell_c}(\cdot,\cdot))\text{, for every $j=1,\cdots, n_i$, $i\in c$},\\
     & \theta_c\equiv \{\beta_c,\sigma^2_c,\alpha_c,\ell_c\}\stackrel{i.i.d}{\sim} G_0\text{, for every $c\in\mathcal{C}$},\\
     & \mathcal{C}\sim p_\lambda(\mathcal{C}\mid \mathcal{S}_n),\\
     & dG_0\equiv\pi_0(\beta,\sigma^2)\pi_1(\alpha)\pi_2(l) d\beta d\sigma^2 d\alpha d l,\\
     & \beta_c\mid\sigma_c^2\sim \mathcal{N}(\mu_0,\sigma_c^2\Lambda_0^{-1}),\\
     & \sigma_c^{-2}\sim \text{Gamma}(a_0,b_0),\\
     & \alpha_c \sim \text{Gamma}(a_1,b_1),\\
     & \ell_c \sim \text{Gamma}(a_2,b_2),
    \end{split}
\end{equation}
where $e_{i}$ is the temporal random effect for location $i$ and
$K_{\sigma_c,\ell_c}$ is the associated squared exponential covariance kernel, defined by, $K_{\sigma_c,\ell_c}(t^{(i)}_k,t^{(i)}_l)=\sigma_{c}^2\exp\{-\frac{1}{2\ell_{c}}(t^{(i)}_k-t^{(i)}_l)^2\}.$ 
To incorporate an MRF structure, we use the prior in \eqref{eq6} for $\mathcal{C}$ by choosing $P$ to be a MFM prior, and setting $l(\theta_c)=1$ and $k(c\mid\mathcal{G}_N)=\exp\{\lambda E_c\}$ for $M$, where $\lambda$ is a tuning parameter and $E_c$ denotes the number of edges among the locations assigned to cluster $c$. Also, $a_0,a_1,a_2,b_0,b_1,b_2$ are hyperparameters in their associated gamma distribution. 
These 
yield a MRF-EPPF prior as defined in Theorem \ref{thm2}. For simplicity, we refer this prior as MRF-MFM for the rest of this paper.  Note that $k(c\mid\mathcal{G}_N)$ satisfies \eqref{eq5}, and the corresponding $k_i(\cdot\mid\mathcal{G}_N)$ 
function coincides with the conditional cost function defined in \citet{zhao2020bayesian}. The partition probability function induced by the MRF-MFM prior, denoted by $p_\lambda(\mathcal{C}\mid \mathcal{G}_N)$, equals to  
\begin{equation}
    \begin{split}
     & p_\lambda(\mathcal{C}\mid\mathcal{G}_N)=\frac{V_N(|\mathcal{C}|)\Pi_{c\in\mathcal{C}}\gamma^{(|c|)}\exp\{\lambda E_c\}}{\sum_{\mathcal{C}'\in\mathcal{P}}V_N(|\mathcal{C}'|)\Pi_{c\in\mathcal{C}'}\gamma^{(|c|)}\exp\{\lambda E_c\}},
    \end{split}
\end{equation}
where $\mathcal{P}$ is the set of all possible partitions of $[N]$. As discussed in \citet{miller2018mixture}, $\gamma$ is the parameter of the symmetric Dirichlet distribution defined in the MFM prior, and $V_N(t)=\sum_{k=1}^\infty\frac{k_{(t)}}{(\gamma k)^{(N)}}p_K(k)$, with $x^{(m)}=\frac{\Gamma(x+m)}{\Gamma(x)}$, $x_{(m)}=\frac{\Gamma(x+1)}{\Gamma(x-m+1)}$, $x^{(0)}= x_{(0)}=1$.

In practice, we let $\lambda\geq 0$, with a larger value of $\lambda$ representing a higher spatial correlation. It can be seen that when $\lambda=0$, $p_\lambda(\mathcal{C}\mid\mathcal{G}_N)$ can recover the partition probability function induced by the MFM prior without any spatial correlation between locations, and when $\lambda\to\infty$, it will degenerate to the Dirac delta function $\delta_{[N]}$. The term $\exp\{\lambda E_c\}$ will change the prior's preference on different partitions, and the prior mass will   concentrate on those partitions with more within-cluster edges. Therefore, $\lambda$ is called \emph{spatial smoothness parameter} in   \citet{zhao2020bayesian}.

As the partition probability function of MFM is an EPPF, the closed form full conditional distribution for our model can be conveniently obtained by Theorem \ref{thm2} in the following lemma. We omit the proof here because it is based on a very similar calculation with that of Theorem 2.1 in \citet{zhao2020bayesian}.
\begin{lemma}\label{lm3}
For model \eqref{eq8}, the conditional distribution of $\theta_i$ given $\theta_{(-i)}$, the induced partition $\mathcal{C}_i$, and distinct value $\{\theta_c\}_{c\in\mathcal{C}_i}$, is proportional to
\begin{equation}\label{eq9}
    \begin{split}
     & \frac{V_N(|\mathcal{C}_i|+1)}{V_N(|\mathcal{C}_i|)}G_0+\sum_{c\in\mathcal{C}_i}\exp\{\lambda\sum_{j\in c\cap \partial{(i)}}\mathbf{1}(\theta_j=\theta_i)\}(|c|+\gamma)\delta_{\theta_c}.
    \end{split}
\end{equation}
\end{lemma}

\section{Theoretical Properties}\label{sec 2.3: Theoretical Properties}
We investigate the asymptotic property for the proposed method and show a clustering consistency result in this section. Note that the asymptotics in our model refers to the situation that the number of spatial locations $N$ is fixed, and the number of observed time points, denoted by $n_i$ for location $i$, goes to infinity. Then our clustering consistency result provides a useful justification for our method in the sense that as we are collecting more data over time for each spatial location, the proposed method will be able to correctly identify the true unknown clustering structure with posterior probability tending to one.  

We first introduce some notations. Let $\mathcal{C}_0$ be the true unknown partition (clustering) structure,  $\mathcal{P}_0=\{\mathcal{C}_0\}$, and $\mathcal{P}$ be the collection of all partitions of $[N]$. Let $\mathcal{P}_1$ be the collection of over-clustering partitions, i.e., $\mathcal{P}_1 = \{\mathcal{C}_1: \mathcal{C}_1 \neq \mathcal{C}_0,~\text{and}~ \forall c'\in\mathcal{C}_1, \exists~ c\in\mathcal{C}_0,\text{s.t.}~ c'\subseteq c\}$.   and $\mathcal{P}_2=\mathcal{P}\setminus (\mathcal{P}_0\cup\mathcal{P}_1)$ be the collection of mis-clsutering partitions. We focus on the model defined in \eqref{eq2}, and denote the response and covariates for location $i$ by $\{Y_i,X_i\}$. Let  $BF_{\mathcal{C},\mathcal{C}_0}=\frac{\Pi_{c\in\mathcal{C}}m(Y_c\mid X_c)}{\Pi_{c\in\mathcal{C}_0}m(Y_c\mid X_c)}$ be the Bayes factor by comparing the regression models given partition $\mathcal{C}$ with the true model $\mathcal{C}_0$, where $m(Y_c\mid X_c)$ is the conditional marginal likelihood of $\{Y_{i},X_{i}\}\text{, for $i\in c$}$. Furthermore, for any partition probability function $p(\mathcal{C})$, we consider its MRF constrained version, modified by the joint cost function introduced in \eqref{eq8}, namely
\begin{equation}\label{eq12}
    \begin{split}
     & p_{\lambda}(\mathcal{C}\mid \mathcal{G}_N)\propto p(\mathcal{C})\Pi_{c\in\mathcal{C}}\exp\{\lambda E_c\}.
    \end{split}
\end{equation}
Define $p_{\max}=\text{max}_{\mathcal{C}\in\mathcal{P}} p(\mathcal{C})$ and let $E_{\max}$ be the total edges among these $N$ locations. Note that both $E_{\max}$ and $p_{\max}$ are finite, since the location number $N$ is finite. Let $n_{\min} = \min\{n_1,\ldots,n_N\}$. We make the following assumptions.
\begin{enumerate}
\item[(A0)] No isolated island: for every $ c\in\mathcal{C}_0$, we assume that $ |\partial{(i)}\cap c|\geq 1$  for every $i\in c$.
\item[(A1)] Model identifiability:  we assume $\theta_c$ is within the support of $G_0$ for every $c\in\mathcal{C}_0$. Moreover, the mixture model \eqref{eq2} is identifiable in the sense that  for any $c\subseteq [N]$, $\Pi_{i\in c}f_{\theta_c}(Y_i\mid X_i)=\Pi_{i\in c}f_{\theta_c'}(Y_i\mid X_i)$ implies $\theta_c=\theta_c'$.
\item[(A2)] Control the mis-clustering partitions: for every $\mathcal{C}\in\mathcal{P}_2$, there exists a sequence of numbers $q_{\mathcal{C}}(n_{\min})$ such that  $BF_{\mathcal{C},\mathcal{C}_0}=o_p(q_{\mathcal{C}}(n_{\min}))$ and  $q_{\mathcal{C}}(n_{\min})\to 0$ as $n_{\min}\to\infty$. 
\item[(A3)] Control the over-clustering partitions: $BF_{\mathcal{C},\mathcal{C}_0}=O_p(1)$ as $n_{\min}\to \infty$, for every $  \mathcal{C}\in\mathcal{P}_1$,
\item[(B3)] Control the over-clustering partitions: $BF_{\mathcal{C},\mathcal{C}_0}\stackrel{p}{\to}0$ as $n_{\min}\to\infty$, for every $\mathcal{C}\in\mathcal{P}_1$.  
\end{enumerate}
Selecting a clustering partition structure can be viewed as a model section problem. Under the Bayesian framework, to correctly identify the true model usually requires the Bayes factor between the true and incorrect models to converge to $0$, which is why Assumptions (A2) and (B3) are needed. In particular, (A2) can be interpreted as, one only needs to find an upper bound $q_{\mathcal{C}}(n_{\min})$ for the contraction rate of the Bayes factor between the true and incorrect models, which is a reasonable assumption since the true model usually has a faster posterior contraction rate than that of an incorrect model if the Bayes Factor is consistent \citep{chib2016bayes}. 
Assumption (A0) and (A3) are alternative replacements for (B3) that allows a weaker rate condition on the Bayes factor between the true and over-clustered models. Assumption (A1) 
is needed for model identifiability purpose, and it is satisfied for many regression problems. 
Let $\Pi ( \cdot \mid \mathcal{G}_N, \{Y_{i},X_{i}\}_{i=1}^N)$ be the posterior distribution given the collected data and the spatial graphic information, we can state the following clustering consistency theorem. 
\begin{theorem}\label{thm3}
Consider model \eqref{eq2} with a prior $p(\mathcal{C})$ being specified in \eqref{eq12}.
Assume that $p(\mathcal{C}_0)>0$, and Assumption (A1), (A2), (B3) hold, then for any $\lambda\geq 0$, we have
\begin{equation}\label{eq11}
    \begin{split}
     & \Pi (\mathcal{C}=\mathcal{C}_0\mid \mathcal{G}_N, \{Y_{i},X_{i}\}_{i=1}^N)\to 1,~~ \text{as} ~n_{\min}\to\infty. 
    \end{split}
\end{equation}
If (A0) and (A3) hold instead of (B3), then there exist a sequence of numbers $\lambda_{n_{\min}}\to\infty$ as $n_{\min}\to\infty$, such that \eqref{eq11} holds. 
\end{theorem}
Our theorem implies that if the Bayes factor is consistent (for definition, see \citet{chib2016bayes}) and the true model contracts at a faster rate than the over-clustered model to the truth, then for any partition probability function that assigns a positive probability to the true partition, the weak consistency of clustering holds. Moreover, if the spatial information is available, then we can achieve the same clustering consistency result even when the true model contracts not slower than the over-fitted model. 

In Theorem \ref{thm4}, we further show that if $f_{\theta_c}(y\mid x)$ is a linear regression model, then the weak consistency of clustering can be achieved under weaker conditions. The proof is given in the Appendix.  
\begin{theorem}\label{thm4}
Consider the following linear regression model,
\begin{equation}\label{eq20}
    \begin{split}
     & Y_{i}(t^{(i)}_j)\mid X_{i}(t^{(i)}_j) \stackrel{ind}{\sim} \mathcal{N}(X_{i}(t^{(i)}_j)\beta_c,\sigma_c^2)\text{, for $j=1,\ldots,n_i$, $i\in c$,}\\
     & \beta_c\mid\sigma^2_c\sim \mathcal{N}(\mu_0,\sigma_c^2\Lambda_0^{-1}), \sigma_c^2\sim \text{IG}(\text{shape}=a_0,\text{rate}=b_0) \text{, for $c\in\mathcal{C}$,}\\
     & \mathcal{C}\sim p_{\lambda}(\mathcal{C}\mid \mathcal{G}_N).
    \end{split}
\end{equation}
Then (A1), (A2), and (B3) hold. As a result,  as $n_{\min}\to\infty$,  $$\Pi(\mathcal{C}=\mathcal{C}_0\mid \mathcal{G}_n, \{Y_{i},X_{i}\}_{i=1}^N)\to 1.$$ 
\end{theorem}
Note that the model specification in \eqref{eq8} is slightly different from the one defined in \eqref{eq20}, because of the Gaussian process structure. Therefore Theorem \ref{thm3}    cannot be directly applied to obtain the weak consistency result for \eqref{eq20}. The main argument in the proof is that for \eqref{eq20}, $\{Y_i,X_i\}_{j=1}^{n_i}$ is exchangeable for every $ i\in [N]$, and the limiting distribution is well defined and unique when $n_i\to\infty$ by Kolmogorov's extension theorem. Then based on the De Finetti's theorem \citep{de1929funzione}, there exists a latent distribution, conditioning on which $\{Y_i(t_j^{(i)}),X_i(t_j^{(i)})\}_{j=1}^{n_i}$ can be treated as independent samples. The remaining task is then to establish Bayes factor consistency and verify assumptions for \eqref{eq20}.

\section{Bayesian Inference}\label{sec 3: Bayesian Inference}
\subsection{Algorithm}\label{sec 3.1: Algorithm}
As discussed in Section \ref{sec 2.1: MRF-PPM}, the Polya urn scheme cannot be applied in our case. As an alternative, we resort to the Algorithm 8 of \citet{neal2000markov} and propose a convenient posterior sampling algorithm. Consider $\phi_c=\{\alpha_c,\ell_c\}$, of which the posterior distribution is intractable. In addition to the notations given in Section \ref{sec 2.2: Model Specification}, we define
\begin{equation}\label{eq25}
    \begin{split}
     & Y_i=[Y_i(t^{(i)}_1),\ldots,Y_i(t^{(i)}_{n_i})]^T, X_i=[X_i(t^{(i)}_1),\ldots,X_i(t^{(i)}_{n_i})]^T, \mathbf{1}_i=[1,\ldots,1]_{n_i\times1}^T\\
     & t^{(i)}=[t^{(i)}_1,\ldots,t^{(i)}_{n_i}]^T, K_i=\exp\{-\frac{1}{2\ell_{c_i}}(t^{(i)}\mathbf{1}_i^T-\mathbf{1}_i{t^{(i)}}^T)^{(2)}\}+\alpha_{c_i}I_{n_i\times n_i}\\
     & \Lambda_c=\sum_{i\in c}X_i^T{K_i}^{-1}X_i+\Lambda_0, \mu_c=\Lambda_c^{-1}(\sum_{i\in c}X_i^T{K_i}^{-1}Y_i+\Lambda_0\mu_0)\\
     & a_c=\frac{\sum_{i\in c}n_i}{2}+a_0, b_c=b_0+\frac{1}{2}(\sum_{i\in c}Y_i^TK_i^{-1}Y_i+\mu_0^T\Lambda_0\mu_0-\mu_c^T\Lambda_c\mu_c)\\
     & \Lambda_{c_i}=X_i^T{K_i}^{-1}X_i+\Lambda_0, \mu_{c_i}=\Lambda_{c_i}^{-1}(X_i^T{K_i}^{-1}Y_i+\Lambda_0\mu_0)\\
     & a_{c_i}=\frac{n_i}{2}+a_0, b_{c_i}=b_0+\frac{1}{2}(Y_i^TK_i^{-1}Y_i+\mu_0^T\Lambda_0\mu_0-\mu_{c_i}^T\Lambda_{c_i}\mu_{c_i}),\\
    \end{split}
\end{equation}
 where $A^{(2)}=A\circ A$, and $\circ$ represents the Hadamard product. We then summarize the details of computation in Algorithm \ref{algo}. 
\begin{algorithm}[htp]\label{algo}
\SetAlgoLined
{\bf Init:}Initial partition: $\mathcal{C}$, and initial cluster parameters: $\{\beta_c,\sigma^2_c,\alpha_c,\ell_c\}_{c\in\mathcal{C}}$.\\
\For{$iter=1,2,\cdots n_{iter}$}{
    Step (1): Update $\{\beta_c,\sigma^2_c,\alpha_c,\ell_c\}_{c\in\mathcal{C}}$ conditioning on $\mathcal{C}$.\\
    \For{$c\in\mathcal{C}$}{
        \For{$j=1,2,\ldots,n_{rep}$}{
            Sample $\{\beta_c,\sigma^2_c\}$ from the full conditional distribution\\
            $p(\beta_c,\sigma^2_c\mid\{Y_i, X_i\}_{i\in c},\alpha_c,\ell_c,\mathcal{C})$\\
            $\propto (\sigma^{-2}_c)^{a_c}\exp\{-\sigma^{-2}_cb_c\}\text{det}(\sigma_c^2\Lambda_c^{-1})^{-\frac{1}{2}}\exp\{-\frac{1}{2}\sigma^{-2}_c(\beta_c-\mu_c)^T\Lambda_c(\beta_c-\mu_c)\}$.\\
            Sample $\{\alpha_c\}$ from the full conditional distribution via\\
            Metropolis-Hastings. A new value for $\alpha_c$, says $\alpha_c^{new}$, is proposed around $\alpha_c$\\
            using normal proposal.\\
            Sample $\{\ell_c\}$ similar to sampling $\{\alpha_c\}$.\\
        }
    }
    Step (2): Update $\mathcal{C}$ conditioning on $\{\beta_c,\sigma^2_c,\alpha_c,\ell_c\}_{c\in\mathcal{C}}$.\\
    \For{$i=1,2,\cdots N$}{
        Remove index $i$ from a $c\in\mathcal{C}$, says $c_i$, denote the resulting partition by $\mathcal{C^*}$.\\
        \eIf{$|c_i|=1$}{
         Sample $m-1$ pairs of $\{\alpha_k,\ell_k\}_{k=1}^{m-1}$ from the base measure\;
         Denote $\{\alpha,\ell\}_m=\{\{\alpha_{c_i},\ell_{c_i}\},\{\alpha_1,\ell_1\},\ldots,\{\alpha_{m-1},\ell_{m-1}\}\}$\;
         }{
         Sample $m$ pairs of $\{\alpha_k,\ell_k\}_{k=1}^{m}$ from the base measure\;
         Denote $\{\alpha,\ell\}_m=\{\{\alpha_1,\ell_1\},\ldots,\{\alpha_{m},\ell_{m}\}\}$\;
        }
        Put $i$ back into a $c\in\mathcal{C^*}$ with probability\\
        $\propto \exp\{\lambda\sum_{j\in c}\mathbf{1}(\theta_j=\theta_i)\}(|c|+\gamma)\times f(Y_i\mid X_i,\theta_c)$,\\
        or create a new cluster with probability\\
        $\propto \frac{1}{m}\frac{V_n(|\mathcal{C}^*|+1)}{V_n(|\mathcal{C}^*|)}\times\int f(Y_i\mid X_i,\beta,\sigma^2,\alpha_k,\ell_k)d(\beta,\sigma^2)$.\\
        with $\theta_{c_i}=\{\beta_{c_i},\sigma^2_{c_i},\alpha_k,\ell_k\}$ being the parameter of this cluster, where\\
        $p(\beta_{c_i},\sigma^2_{c_i}\mid\{Y_i, X_i\}_{i\in c},\alpha_k,\ell_k,\mathcal{C})$\\
        $\propto (\sigma^{-2}_{c_i})^{a_{c_i}}\exp\{-\sigma^{-2}_{c_i}b_{c_i}\}(\sigma_{c_i}^2\Lambda_{c_i}^{-1})^{-\frac{1}{2}}\exp\{-\frac{1}{2}\sigma^{-2}_{c_i}(\beta_{c_i}-\mu_{c_i})^T\Lambda_{c_i}(\beta_{c_i}-\mu_{c_i})\}$.\\
        Let $\mathcal{C}$ be the resulting partition.
    }
}
 \caption{posterior sampling scheme for MRF-PPM.}
\end{algorithm}
Details about prior specification are given at the end of Section \ref{sec 4.1: Sim Setting}. In practice, we use the normal proposal with standard error $.01$, and let $n_{rep}=30$, in order to sample from the high-density region.
\subsection{Selection of \texorpdfstring{$\lambda$}{Lg}}\label{sec 3.2: Selection}
The spatial smoothness parameter $\lambda$ plays an important role in our model and how to decide its value in practice can be viewed as a model selection problem. It is common to consider using the marginal likelihood function as a selection criteria, whose value, however, is intractable for many Bayesian complex models including the one we discussed in this paper. Several posterior sampling based approaches have been proposed in the literature, such as logarithm of the
Pseudo-marginal likelihood (LPML) \citep{lewis2014posterior} and marginal likelihood computed by Harmonic Mean \citep{newton1994approximate}. However, these methods usually suffer from the Pseudo-Bias issue \citep{lenk2009simulation}, which tends to prefer the model with higher complexity. 
Moreover, in our Algorithm \ref{algo}, it is difficult to split an observation (location in our setting) from its current cluster and create a new cluster by itself after the algorithm is trapped into a local maxima. This happens more frequently if a vague prior is used, which makes it impossible for the posterior sampling based approaches to fully explore the entire posterior space. Sequential Importance Sampling Method \citep{basu2003marginal} is another popular approach for marginal likelihood estimation, but it cannot be applied to MRF-PPM because the Polya urn scheme is not available for MRF-PPM. 

Our solution is to consider a prior sampling based approach for estimating  the marginal likelihood as follows,  
\begin{equation}\label{eq26}
    \begin{split}
     &  \widehat{m}(Y\mid X)=\frac{1}{M-M'}\sum_{k=(M'+1)}^{M}\Pi_{c\in\mathcal{C}_{(k)}}f(Y_c\mid X_c,\alpha_k,\ell_k),
    \end{split}
\end{equation}
where $\mathcal{C}_{(k)}$,$\alpha_k$ and $\ell_k$ are the associated parameters sampled from the partition probability function at the $k$-th iteration, and $f(Y_c\mid X_c,\alpha_k,\ell_k)$ is the likelihood function by integrating out $(\beta,\sigma^2)$ on its prior. More specifically, $\mathcal{C}_{(k)}$ at each iteration is sampled via the Gibbs sampler defined by \eqref{eq9}. To account for the potential high level of variation in the prior sampling estimate, we follow the suggestion in \citet{basu2003marginal} by letting $\mathcal{C}_{(0)}$ (the initial partition) equal to the last sample from Algorithm \ref{algo}, with $n_{iter}=1000$ and the first 500 iterations be burn-in iterations, $M=10^{6}$, $M'=10^{4}$ (burn-in procedure) and set the same random seed for different $\lambda$ value. In both simulation and real data analysis, we choose $\lambda$ from $\{0, 0.1,\ldots, 1\}$ and find out that this range works quite well since the selected optimal $\lambda$ is always inside $(0,1)$ in the results.

\section{Simulation}\label{sec 4: Simulation}
In this section, we study the empirical performance of our  MRF-MFM model and compare with four approaches, including two Bayesian methods without accounting for spatial correlation, namely, the Dirichlet Process (DP) and MFM, and two frequentist methods in  \citet{lin2012estimation} and \citet{su2016identifying}. In the numerical analysis, Dahl's method \citep{dahl2006model} will be used to summarize the posterior samples and obtain a deterministic result for both cluster assignment and cluster-wise parameter. Rand index (RI;  \citet{rand1971objective}) will be adopted as a metric to evaluate the discrepancy between different partitions. All computations  were performed on 10 computing servers. Each server has 94.24GB RAM, 24 processing cores, and operates at 3.33 GHz. We distributed our simulation task to 100 workers (10 cores for each server), and it  took approximately 20 hours to finish each simulation scenario with 100 Monte Carlo replications including the tuning procedure.

\subsection{Simulation Setting}\label{sec 4.1: Sim Setting}
\begin{figure}[htp]
\minipage{0.5\textwidth}
  \includegraphics[width=\linewidth]{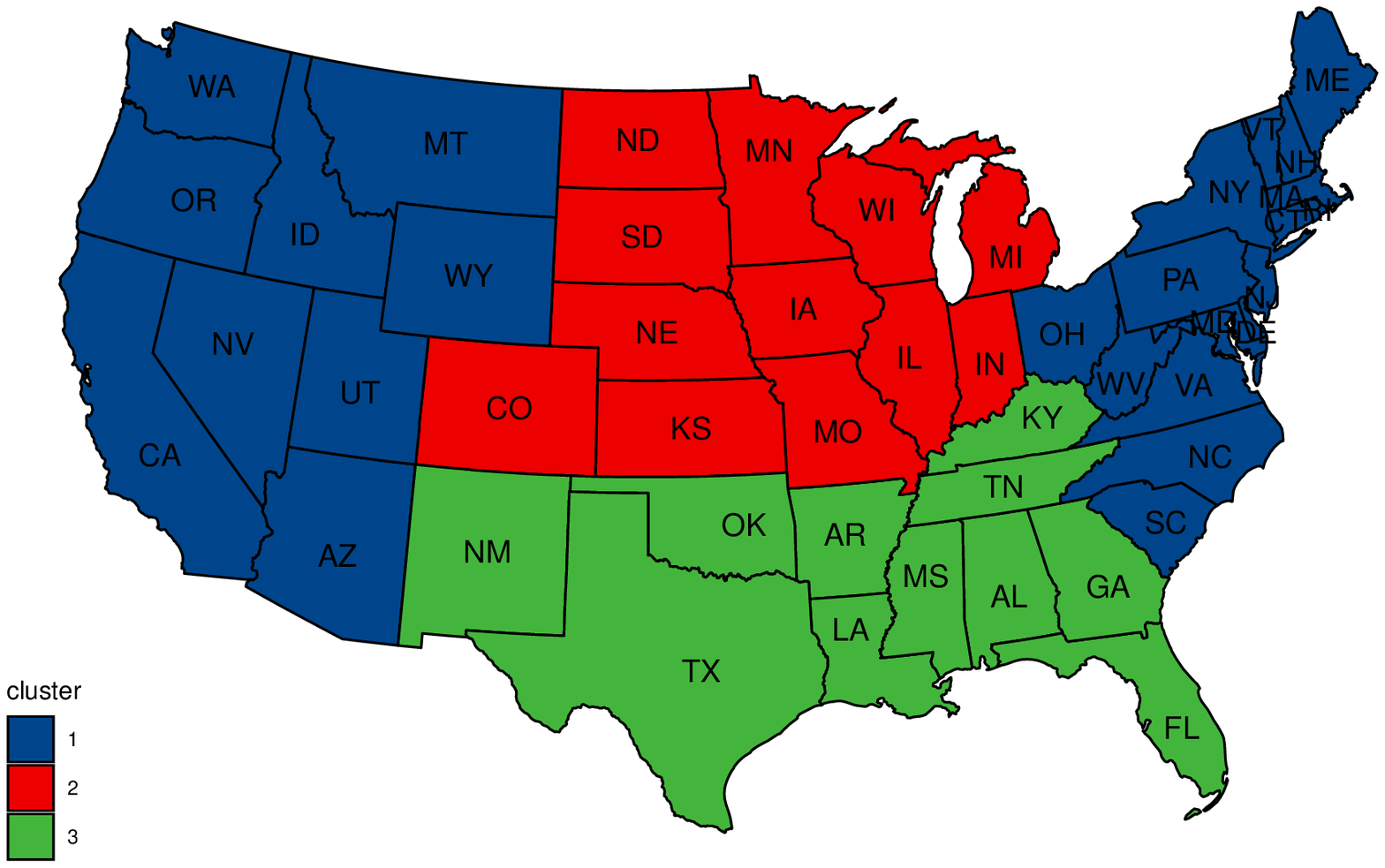}
\endminipage\hfill
\minipage{0.5\textwidth}
  \includegraphics[width=\linewidth]{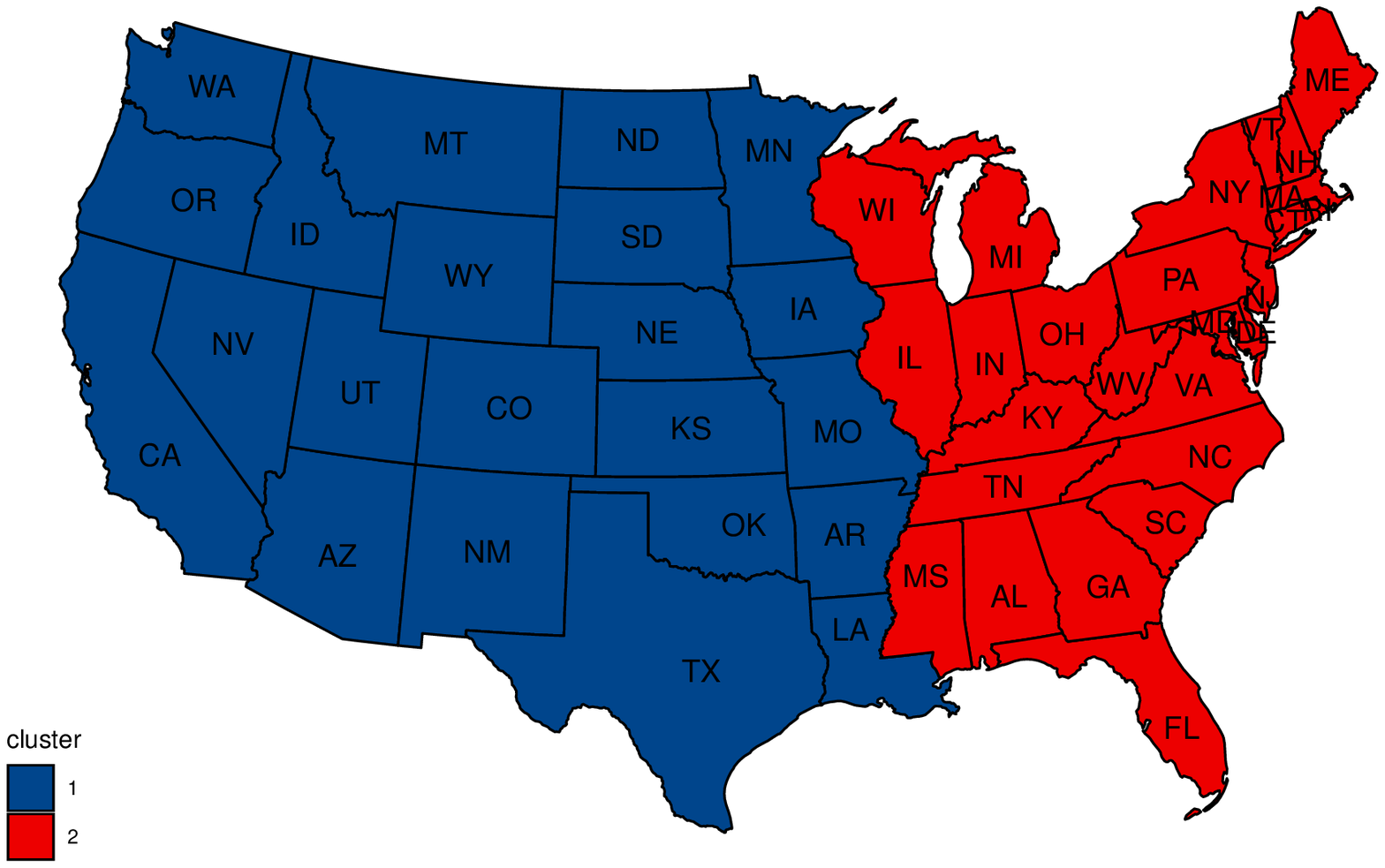}
\endminipage\hfill
\caption {\label{fig:1} Simulation partition scenarios 1 and 2}
\end{figure}

For simulation data generation, we consider two partition scenarios (see Figure \ref{fig:1}) with 48 states in the United States, excluding Hawaii, Alaska and the District of Columbia. Both partition scenarios indicate strong spatial correlation, as most individual units assigned to the same cluster are spatially contiguous. The main difference between these two partition settings is that the first partition is more complex since it allows two spatially non-contiguous blocks belonging to the same cluster. 

For each partition scenario, we generate data from the following model,
\begin{equation}\label{eq27}
    \begin{split}
     & Y_{i}(t^{(i)}_j)\mid f_{i}(t^{(i)}_j),X_{i}(t^{(i)}_j) \stackrel{\text{ind}}{\sim} \mathcal{N}(X_{i}(t^{(i)}_j)\beta_{c_i}+f_{i}(t),\sigma_{c_i}^2\cdot \alpha_{c_i}),\\
     & f_{i}(t^{(i)}_j)\sim \mathcal{N}(0,K_{\sigma_{c_i},\ell_{c_i}}(\cdot,\cdot))\text{, for $j=1,\cdots, n_i$, $i\in c$},
    \end{split}
\end{equation}
where $K_{\sigma^2_c,\ell_c}$ is the squared exponential kernel, as defined in \eqref{eq8}, $X_i=[\mathbf{1},\mathbf{x}_i]$, with each entry in $\mathbf{x}_i$ independently sampled from $\text{Unif}(-5,5)$ and $\{t_i\}_{i=1}^{20}$ equally spaced in $[-1,1]$. We consider eight data generating processes (DGP) with different cluster-wise parameters, detailed as follows,

DGP1. $\beta_1=(\epsilon_1,1+\epsilon_2)^T$, $\beta_2=(28,1)^T$, $\beta_3=(-28,1)^T$; $\sigma^2_1=\sigma^2_2=\sigma^2_3=36$; $\alpha_1=\alpha_2=\alpha_3=0.1$; $\ell_1=\ell_2=\ell_3=10$, $\epsilon_i\sim \mathcal{N}(0,0.1)$,

DGP2. $\beta_1=(\epsilon_1,1+\epsilon_2)^T$, $\beta_2=(28,1)^T$, $\beta_3=(-28,1)^T$; $\sigma^2_1=\sigma^2_2=\sigma^2_3=36$; $\alpha_1=\alpha_2=\alpha_3=0.1$; $\ell_1=\ell_2=\ell_3=10$, $\epsilon_i\sim \mathcal{N}(0,0.01)$,

DGP3. $\beta_1=(14+\epsilon_1,1+\epsilon_2)^T$, $\beta_2=(-14,1)^T$; $\sigma^2_1=\sigma^2_2=36$; $\alpha_1=\alpha_2=0.1$; $\ell_1=\ell_2=10$, $\epsilon_i\sim \mathcal{N}(0,0.1)$,

DGP4. $\beta_1=(14+\epsilon_1,1+\epsilon_2)^T$, $\beta_2=(-14,1)^T$; $\sigma^2_1=\sigma^2_2=36$; $\alpha_1=\alpha_2=0.1$; $\ell_1=\ell_2=10$, $\epsilon_i\sim \mathcal{N}(0,0.01)$,

DGP5. $\beta_1=(\epsilon_1,5+\epsilon_2)^T$, $\beta_2=(-20,4)^T$, $\beta_3=(20,6)^T$; $\sigma^2_1=\sigma^2_2=\sigma^2_3=36$; $\alpha_1=\alpha_2=\alpha_3=0.1$; $\ell_1=\ell_2=\ell_3=10$, $\epsilon_i\sim \mathcal{N}(0,0.1)$,

DGP6. $\beta_1=(\epsilon_1,5+\epsilon_2)^T$, $\beta_2=(-20,4)^T$, $\beta_3=(20,6)^T$; $\sigma^2_1=\sigma^2_2=\sigma^2_3=36$; $\alpha_1=\alpha_2=\alpha_3=0.1$; $\ell_1=\ell_2=\ell_3=10$, $\epsilon_i\sim \mathcal{N}(0,0.01)$,

DGP7. $\beta_1=(10+\epsilon_1,5+\epsilon_2)^T$, $\beta_2=(-10,4)^T$; $\sigma^2_1=\sigma^2_2=36$; $\alpha_1=\alpha_2=0.1$; $\ell_1=\ell_2=10$, $\epsilon_i\sim \mathcal{N}(0,0.1)$,

DGP8. $\beta_1=(10+\epsilon_1,5+\epsilon_2)^T$, $\beta_2=(-10,4)^T$; $\sigma^2_1=\sigma^2_2=36$; $\alpha_1=\alpha_2=0.1$; $\ell_1=\ell_2=10$, $\epsilon_i\sim \mathcal{N}(0,0.01)$,

Among them, DGPs 1, 2, 5, 6 are for the partition scenario 1, and the other four are for the scenario 2. Because of different variance magnitude of the random error, we call DGP 1, 3, 5, 7 the strong noise design, and DGP 2, 4, 6, 8 the weak noise design.

In both simulation and real data analysis, we set $\gamma=1$ and $p_K(\cdot)=\frac{10^{k-1}e^{10}}{(k-1)!}$, which corresponds to a $\text{Poisson}(10)$ distribution truncated to positive integers. Empirically, the MFM prior with this parameter setting tends to slightly over-cluster locations, with cluster size evenly distributed. We set the hyper-parameters as $\mu_0=\mathbf{0}_{p \times 1}$, $\Lambda_0=10^{-6}\cdot \text{I}_{p\times p}$, $a_0=0.1$, $b_0=1$. Throughout the numerical studies, we find that the results are not sensitive to the choice of those values. In addition, we set $a_1=a_2=2$ and $b_1=b_2=1$ to encourage $\alpha$ and $\ell$ concentrate around relatively small values.
\subsection{Results}\label{sec 4.2: Sim Results}
We conduct 100 Monte Carlo replications for each of the eight DGPs, and summarize the mean and the median of the rand index obtained by comparing the partition from Dahl's estimate with the ground truth. For the Bayesian methods without spatial smoothness, we let the concentration parameter $\alpha=1$ for DP, and set the parameters of MFM to be the same with those of MRF-MFM in Section \ref{sec 4.1: Sim Setting}. In each Monte Carlo replication, 1000 MCMC iterations are implemented, with the first 500 iterations discarded as the burn-in. For the method in \citet{lin2012estimation}, we follow the default setting of their code, which assumes the number of clusters $|\mathcal{C}|$ is within $\{2,3,4\}$, and selects $|\mathcal{C}|$ based on BIC. The partition is determined following Conditional K-means (CK-means) 
criteria, which is pointed out in their paper to be more robust than the other methods when $n_{\min}$ is small. For the method in \citet{su2016identifying}, we use penalized least squares approach (PLS) to fit the model, and follow the default setting of their code, which assumes $|\mathcal{C}|$ is within $\{1,\ldots,5\}$.  Since both frequentist models can only discriminate latent groups when they have different slopes, they will only be implemented for DGPs 5--8. 
\begin{table}[htp]
    \centering
        \caption{Median (Mean) of random index over 100 Monte Carlo replications for our MRF-PPM method and four competing methods, MFM, DP, Ck-means \citep{lin2012estimation}, and PLS \citep{su2016identifying}.}
    \label{tab:1}
   \begin{tabular}{cccccc}
    \toprule 
DGP  & MRF-MFM & MFM & DP & CK-means & PLS\\
    \midrule 
1& \textbf{0.973} (\textbf{0.924}) & 0.879 (0.886) & 0.909 (0.894) & - & -\\
2& \textbf{1.000} (\textbf{0.929}) & 0.968 (0.928) & 0.969 (0.926) & - & -\\
3& \textbf{1.000} (\textbf{0.981}) & 0.915 (0.917) & 0.938 (0.934) & - & -\\
4& \textbf{1.000} (\textbf{0.990}) & 0.979 (0.954) & 1.000 (0.966) & - & -\\
    \midrule
5& \textbf{0.914} (\textbf{0.916}) & 0.889 (0.901) & 0.911 (0.909) & 0.835 (0.837) & 0.373 (0.373)\\
6& \textbf{1.000} (\textbf{0.982}) & 1.000 (0.975) & 1.000 (0.979) & 0.969 (0.973) & 0.373 (0.373)\\
7& \textbf{0.949} (\textbf{0.934}) & 0.917 (0.915) & 0.936 (0.929) & 0.880 (0.881) & 0.493 (0.493)\\
8& \textbf{1.000} (\textbf{0.984}) & 1.000 (0.970) & 1.000 (0.975) & 0.880 (0.880) & 0.493 (0.493)\\
  \bottomrule
    \end{tabular}
\end{table}

\begin{figure}[htp]
\minipage{1\textwidth}
  \includegraphics[width=\linewidth,height=7cm]{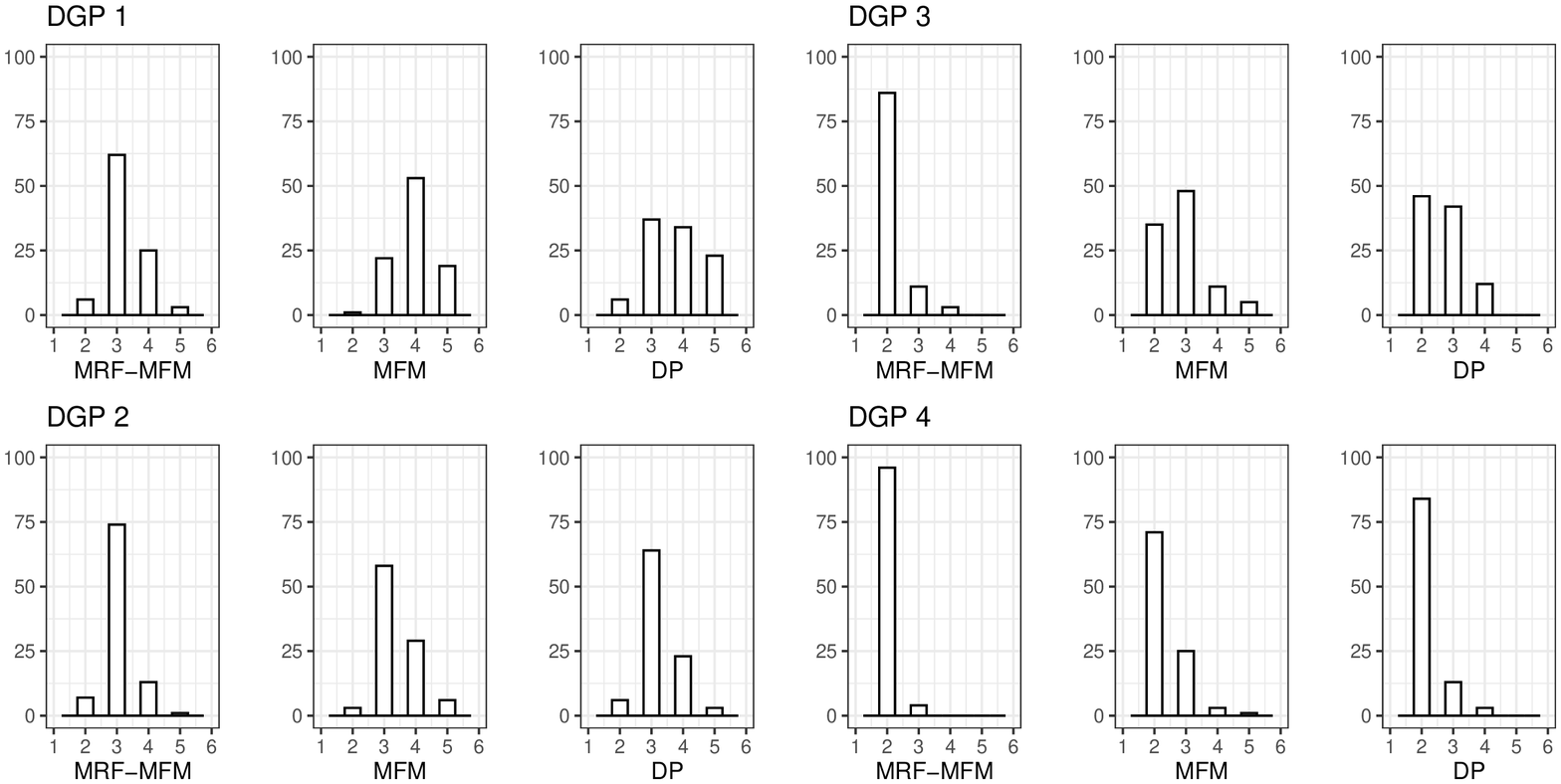}
\endminipage\hfill
\newline
\minipage{1\textwidth}
  \includegraphics[width=\linewidth,height=7cm]{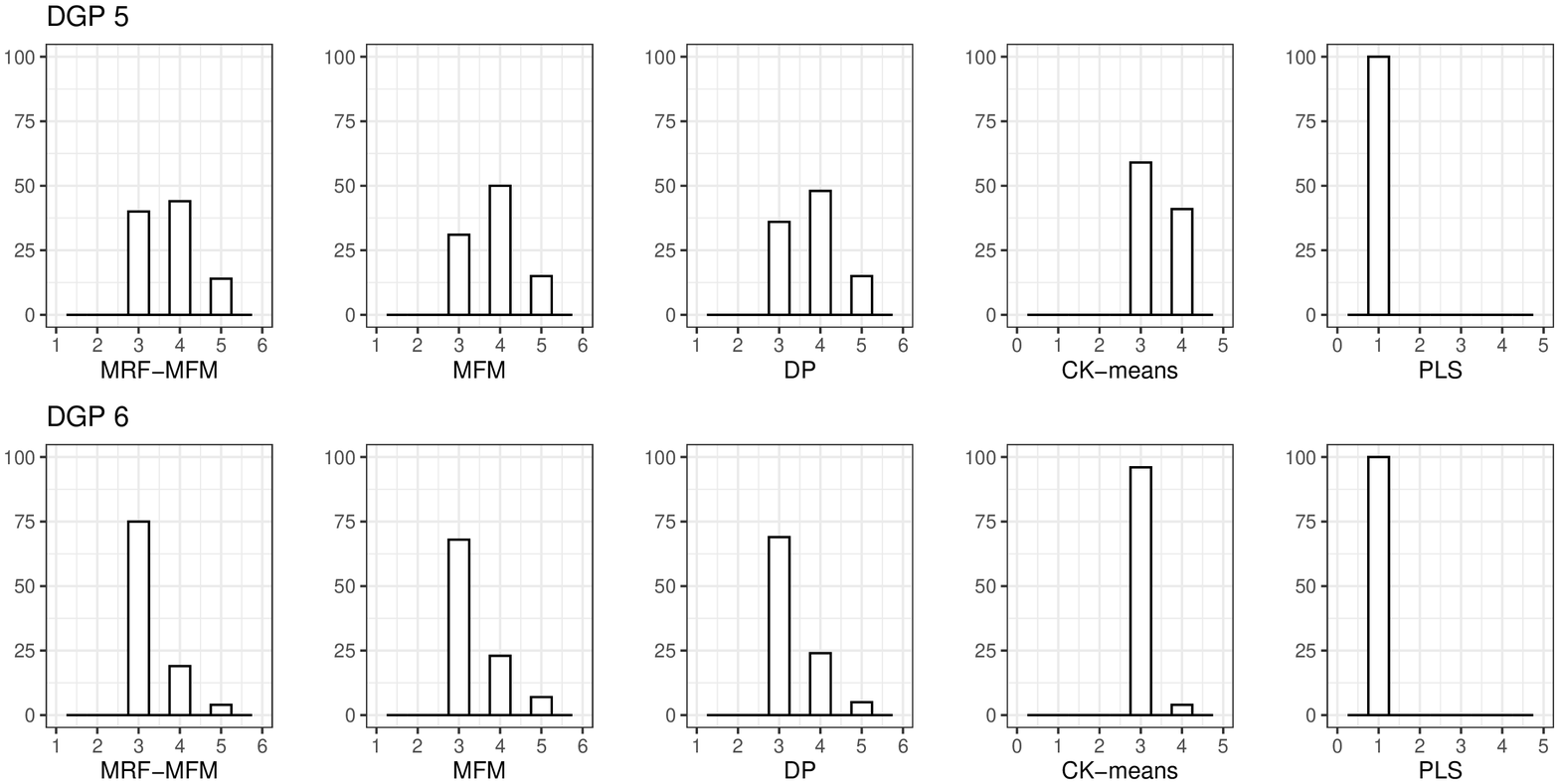}
\endminipage\hfill
\newline
\minipage{1\textwidth}
  \includegraphics[width=\linewidth,height=7cm]{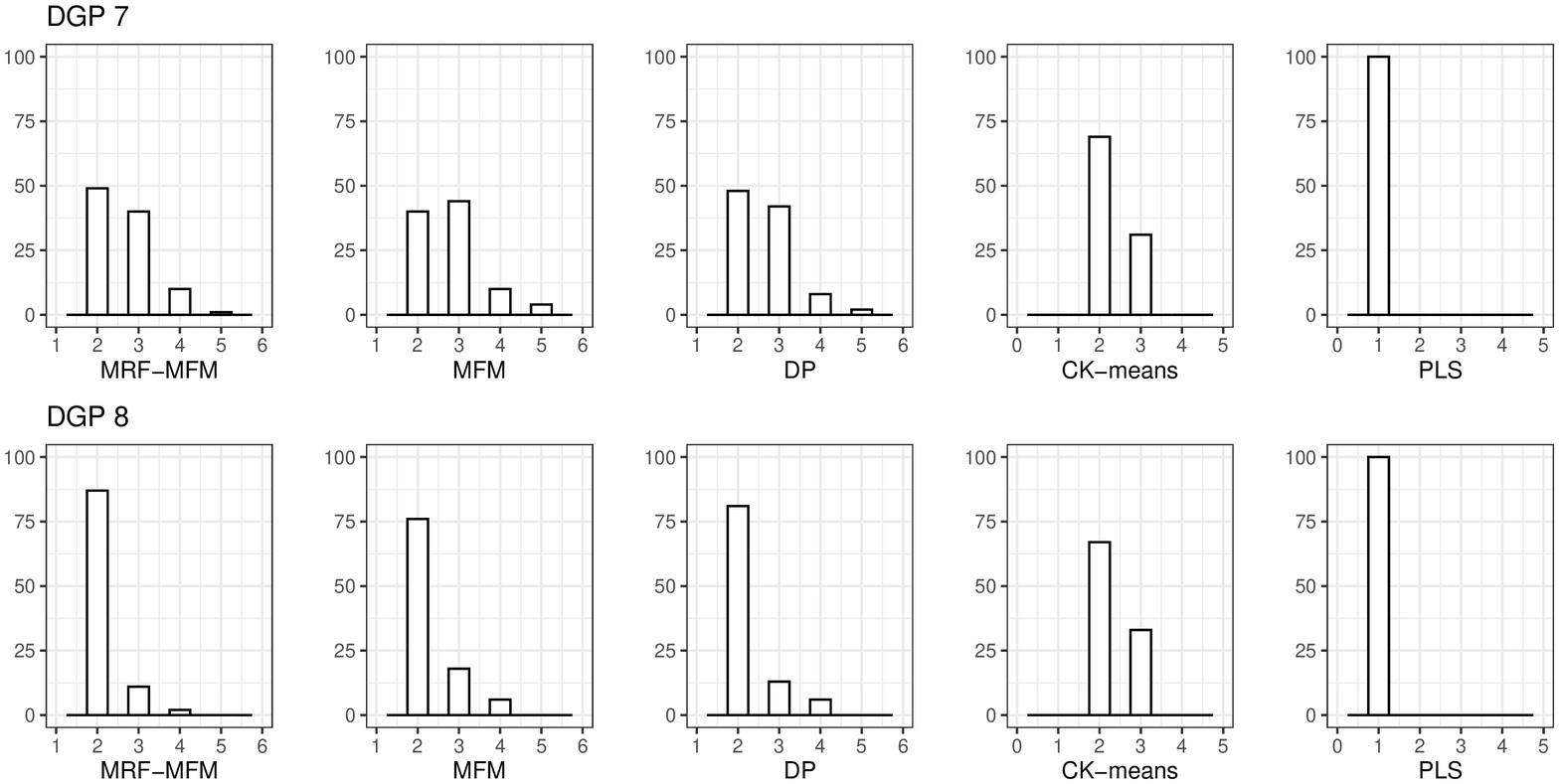}
\endminipage\hfill
\caption {\label{fig:2} Histograms for selected number of clusters by MRF-MFM and four competing methods}
\end{figure}

The simulation results are summarized in Table \ref{tab:1} and Figure \ref{fig:2}. In Table \ref{tab:1}, we find that the proposed MRF-MRF performs uniformly better than the other four methods in terms of a higher value in RI under all scenarios, which confirms the benefit of appropriately incorporating the spatial correlation across different locations. All three Bayesian methods provide a more accurate clustering partition result than that of the two frequentist methods for the reason that those Bayesian methods correctly specify the covariance structure. In general, when the noise level is high (DGPs 1,3,5,7) or the true partition structure becomes more complex (DGPs 1,2,5,6), the RI becomes lower in the table as expected. 

We also compare the CK-means with the PLS method, and find that PLS in general cannot accurately identify the latent group structure in this simulation because the generated data has a strong temporal correlation at each location due to large $\ell$ and small $\alpha$ values. 
This creates trouble for PLS method \citep{su2016identifying}, where they use a z-transformation on both $Y$ and $\mathbf{x}$ and the estimation of slope becomes equivalent to estimating the correlation coefficient. As a consequence, the difference in slopes between clusters cannot be fully captured by their method as the correlation coefficient is close to 1 for all clusters due to high serial correlation. On the other hand, the CK-means method is more robust since it is distance-based. Similar findings are observed in Figure \ref{fig:2}, where we show the histograms for the number of clusters. In general, MRF-MFM has an excellent performance in terms of selecting the correct number of clusters, which is $3$ in our case, for all scenarios. When the partition structure is complex (e.g., DGPs 1,2,5,6). both the DP and MFM tend to overestimate the number of clusters as expected because they do not account for the spatial correlation among locations.

\section{Real Data Analysis}\label{sec 5: Real Data}
In this section, we present two real data applications to demonstrate the use of our proposed methodology. In the data analysis, we choose the spatial smoothness parameter $\lambda$ from the candidate set $\{0,0.1,\ldots,1\}$ based on the criteria described in Section \ref{sec 3.2: Selection}. 

\subsection{Precipitation Data Analysis}\label{sec 5.1: Precipitation}
We first consider the annual precipitation and average temperature data available at \url{https://www.ncdc.noaa.gov/cag/statewide/mapping/110/pcp/201812/12/value} collected by 48 states (exclude Washington, D.C., Alaska and Hawaii) from 2000 to 2019. The main goal is to study the relationship between the annual precipitation and the average  temperature and understand its heterogeneity over different states. It is well known that the precipitation is strongly associated with the convection, which is influenced by the topography \citep{parsons1983relationship}. To account for the spatial heterogeneity, we apply the model in \eqref{eq8}, treat each state as a spatial location $i$, rescale the years from 2000 to 2019 onto equally spaced points between $[-1,1]$, and let $Y_i$ be a 20 by 1 response vector of the annual precipitation and $X_i$ be a 20 by 2 matrix which includes an intercept term and the average temperature as another covariate.

Based on the estimated marginal likelihood, we find the optimal value for $\lambda$ is $\lambda=0.1$. We run 10, 000 MCMC iterations and discard the first 5, 000 as burn-in. The final partition is obtained by Dahl's method. The average rand index between the reporting partition and the 100 replications is 0.9362, which indicates that the final partition is representative. 
\begin{figure}[htp]
\minipage{0.5\textwidth}
  \includegraphics[width=\linewidth]{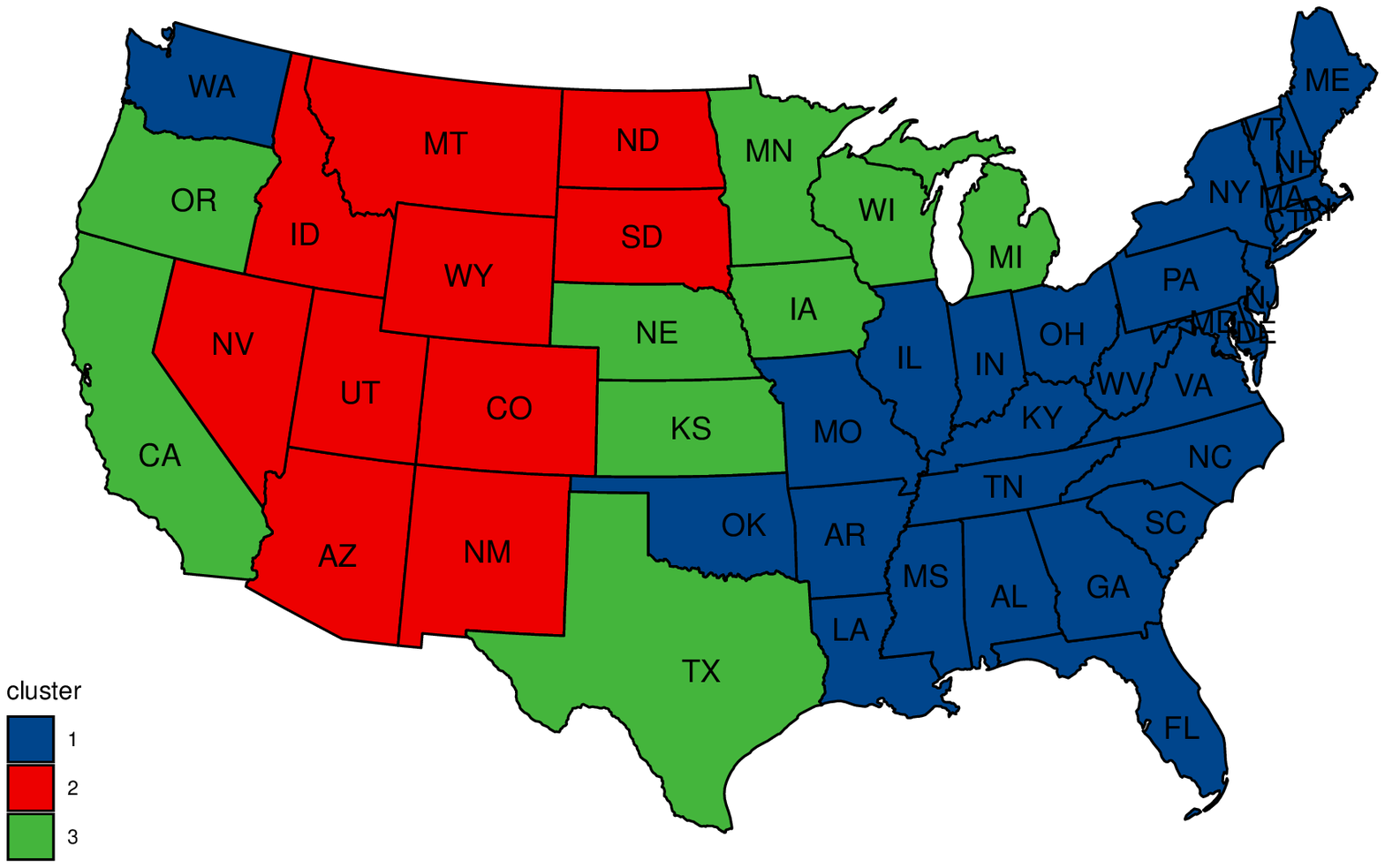}
\endminipage\hfill
\minipage{0.5\textwidth}
  \includegraphics[width=\linewidth]{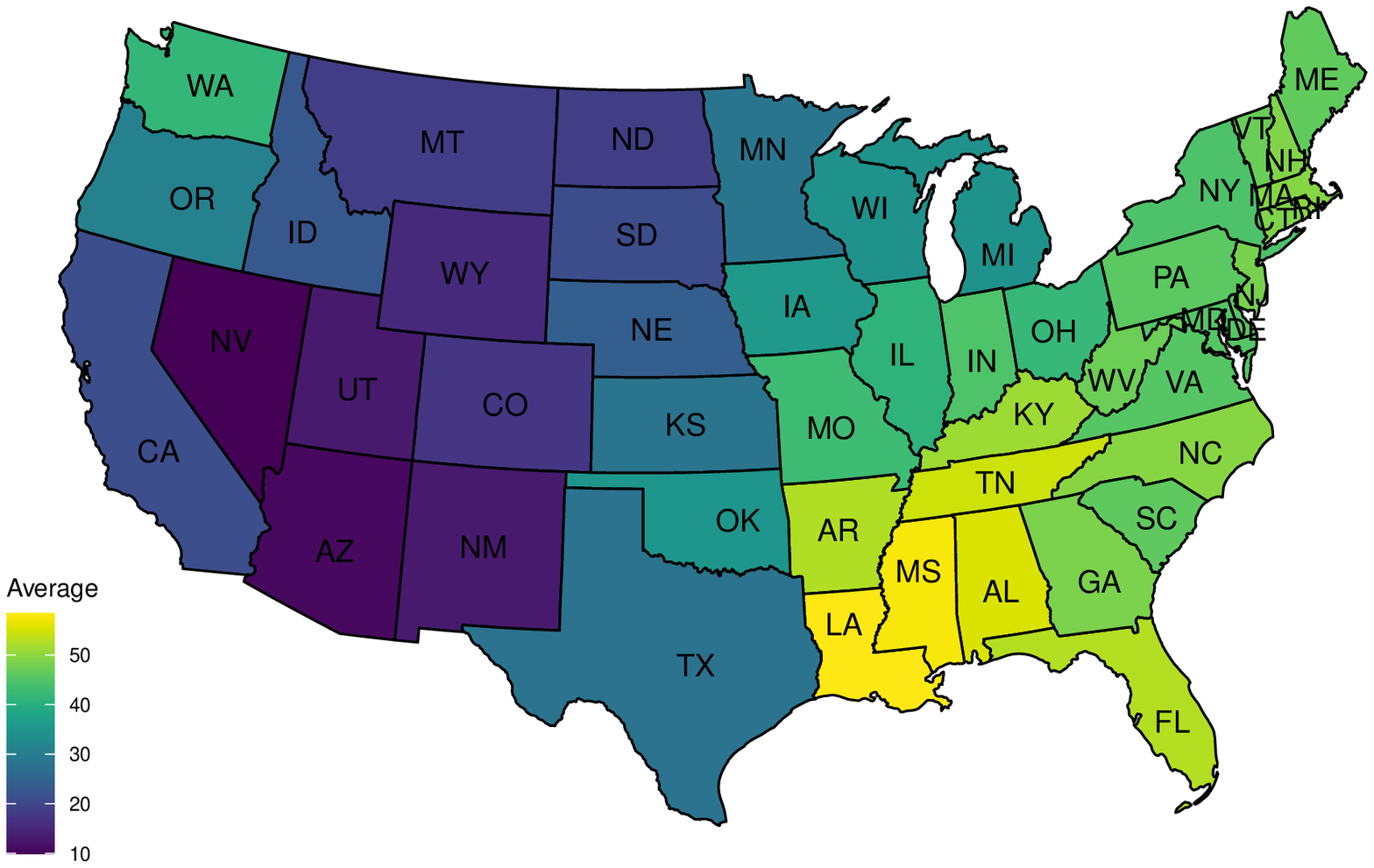}
\endminipage\hfill
\newline
\minipage{0.5\textwidth}
  \includegraphics[width=\linewidth]{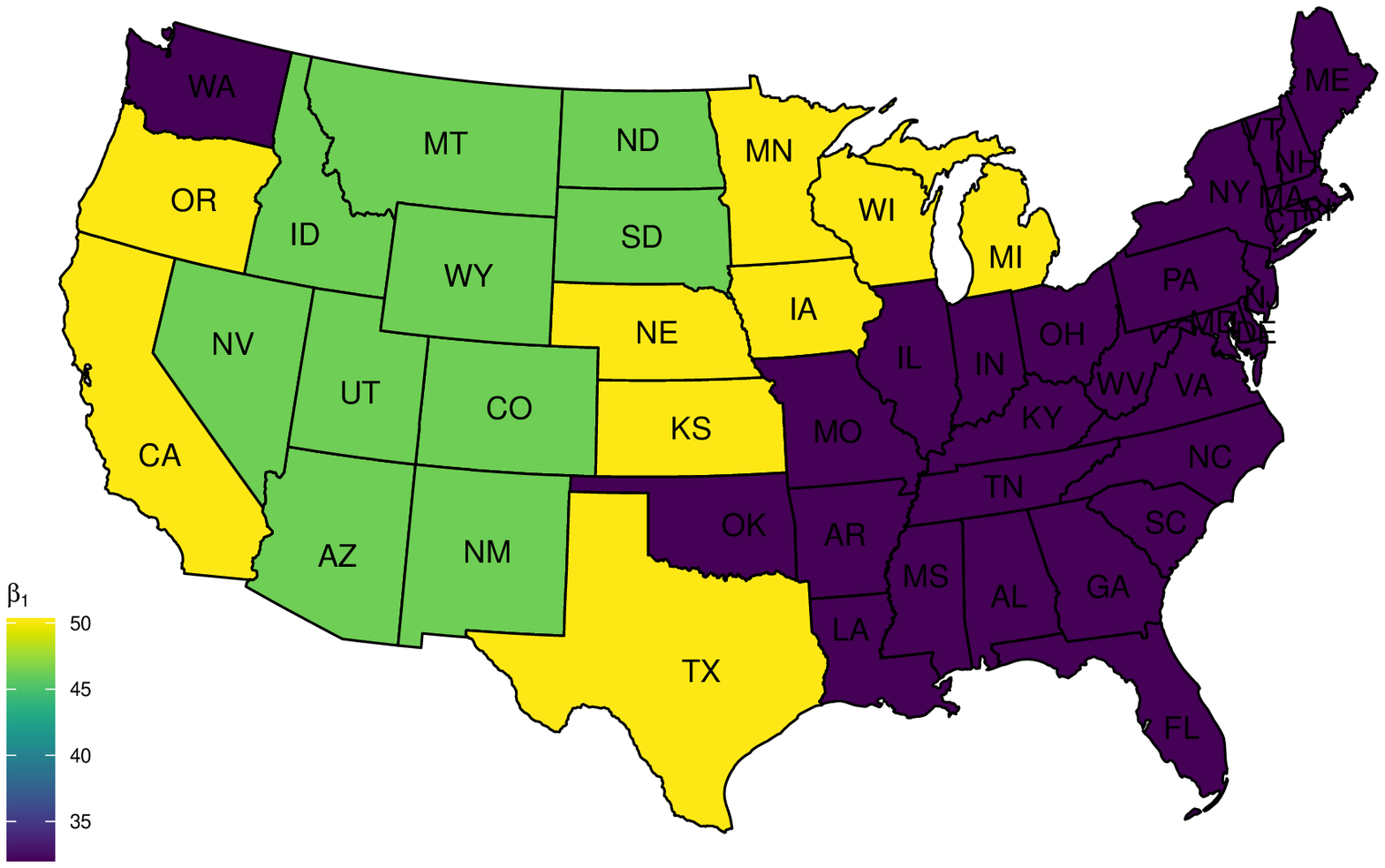}
\endminipage\hfill
\minipage{0.5\textwidth}
  \includegraphics[width=\linewidth]{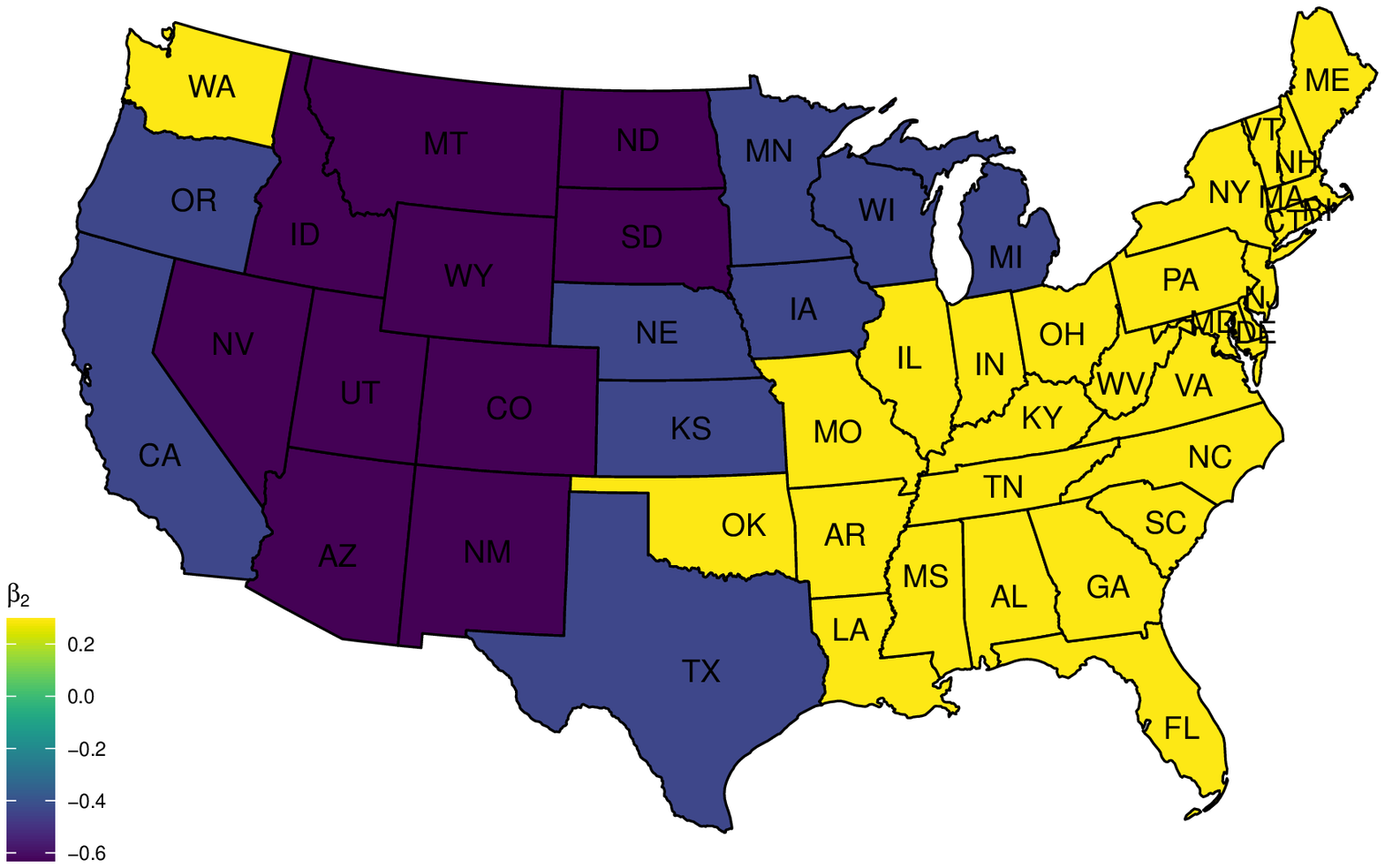}
\endminipage\hfill
\caption {\label{fig:3} Top left: The estimated partition; Top right: The average annual precipitation map;  Bottom left: Estimated intercept; Bottom right: Estimated slope for the annual temperature}
\end{figure}

\begin{table}[htp]
    \centering
        \caption{\label{tab:2} Cluster-wise parameter estimates for the precipitation data}
   \begin{tabular}{cccccc}
    \toprule 
Cluster & $\hat{\beta}_{\text{intercept}}$ & $\hat{\beta}_{\text{tempreature}}$ & $\hat{\sigma^2}$ & $\hat{\ell}$ & $\hat{\alpha}$\\
    \midrule 
1& 32.030 & 0.303 & 13.434 & 1.776 & 3.699\\
2& 46.353 & -0.631 & 3.916 & 5.436 & 1.340\\
3& 50.379 & -0.435 & 17.709 & 5.357 & 1.055\\
    \bottomrule
    \end{tabular}
\end{table}
 
We summarize the obtained partition estimate in Figure \ref{fig:3}. We find that the estimated partition (top left) in general matches the pattern observed in the average annual precipitation map (top right) quite well. More specifically, the first cluster (in blue) contains most states with the climate type of humid continental and humid subtropical, which usually receive plenty of rainfall annually. The climate types of most states in the second cluster (in red), on the other hand, are desert and semi-arid, which naturally associates with a low level of  precipitation. 

In Table \ref{tab:2}, we summarize the estimated regression parameters for each of the obtained three clusters. The  results clearly demonstrate a high level of heterogeneity in both regression coefficients and the variance parameter over three clusters, which again highlights the benefit and  necessity of considering heterogeneity for spatial panel data. 
Scientifically, the mechanism of precipitation is a complex system, and it is known to be more relevant to some other factors, such as the vertical thermal gradient and wind speed. Therefore, one can observe that for the first and third cluster, $\hat{\sigma}^2$ is considerably large, which manifests our statement that there may be some other latent confounders which are not accounted for in our model. 

In our study, we interpret the predictor ``the average annual temperature" as a hybrid indicator, e.g., for the second cluster, the annual temperature seems to  indicate the aridness in the sense that a high level of aridness, which is usually implied by a higher annual temperature, usually leads to less annual precipitation.

\subsection{Median Household Income Data Analysis}\label{sec 5.2: Median Income}
Next we analyze a California State county-level household income dataset available at  \url{https://www.countyhealthrankings.org/app/}). The data consists of annual measurements of median household income, total gross domestic product (GDP) and the unemployment rate between the year of 2011 and 2018. Our interest here is to conduct a regression analysis of the median income on GDP and unemployment rate and study the heterogeneity pattern in the regression parameters over different counties. Before applying our method, we did a logit transformation on the unemployment rate, and a z-transform on the median income and the GDP. 

\begin{figure}[htp]
\minipage{0.5\textwidth}
  \includegraphics[width=\linewidth]{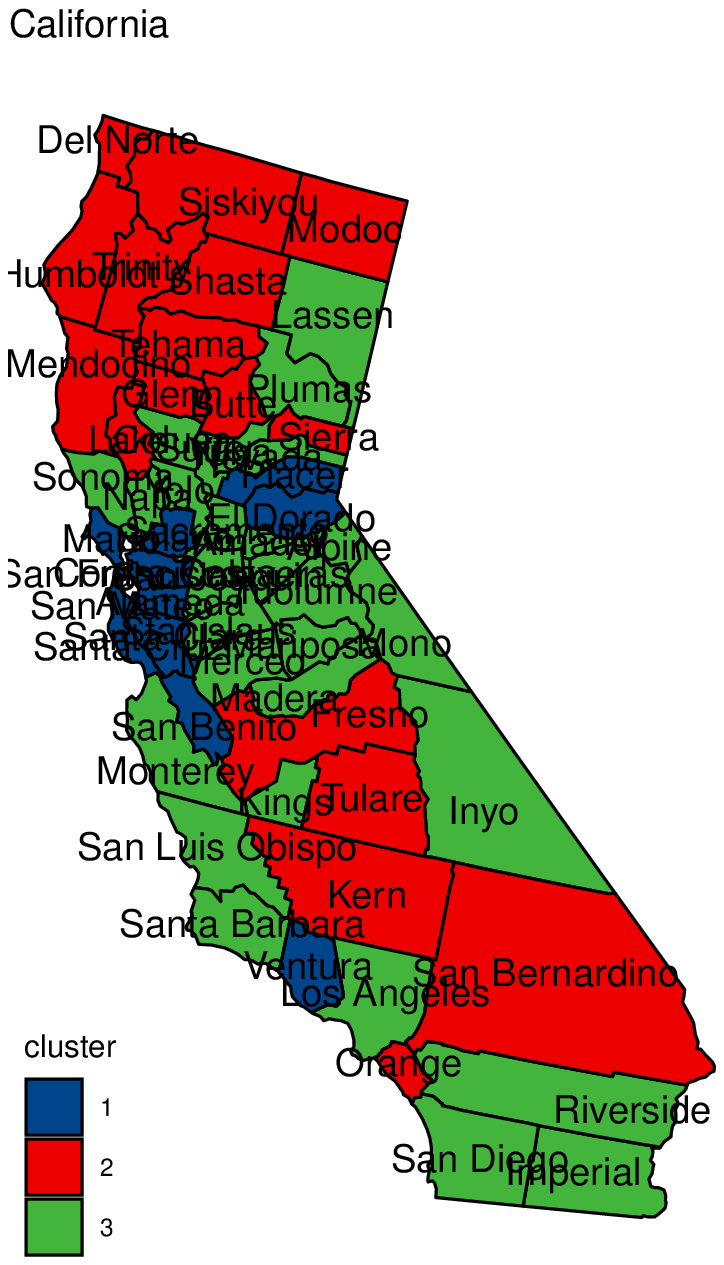}
\endminipage\hfill
\minipage{0.5\textwidth}
  \includegraphics[width=\linewidth]{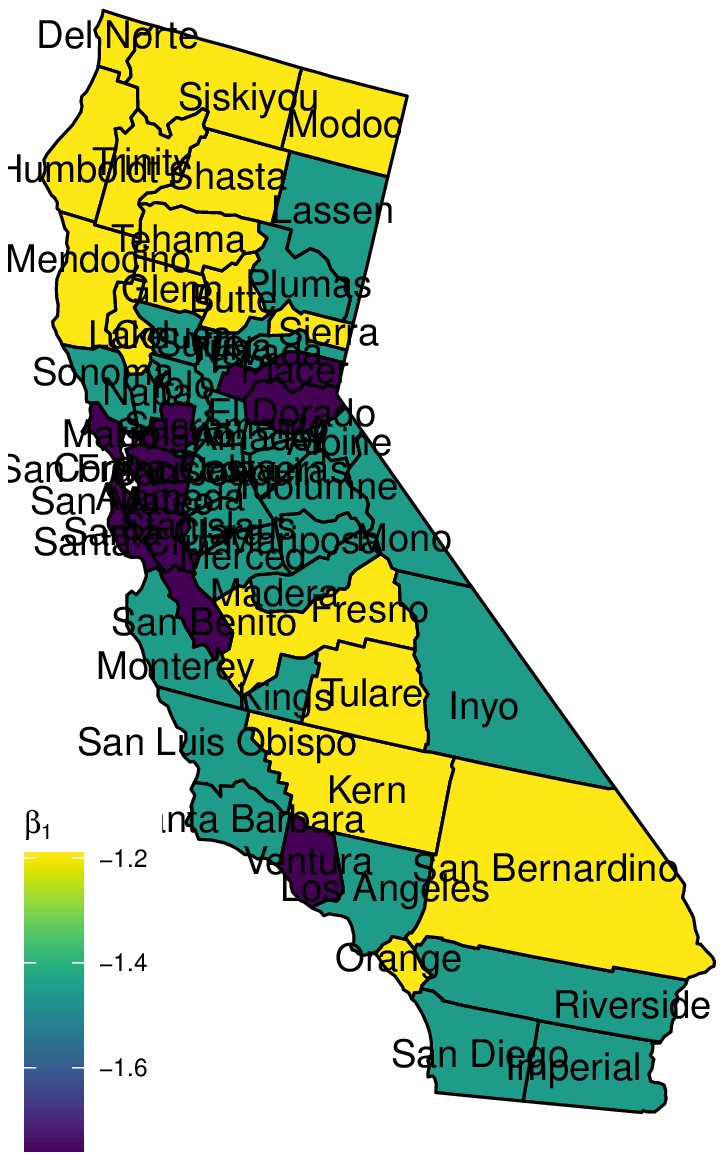}
\endminipage\hfill
\newline
\minipage{0.4\textwidth}
  \includegraphics[width=\linewidth]{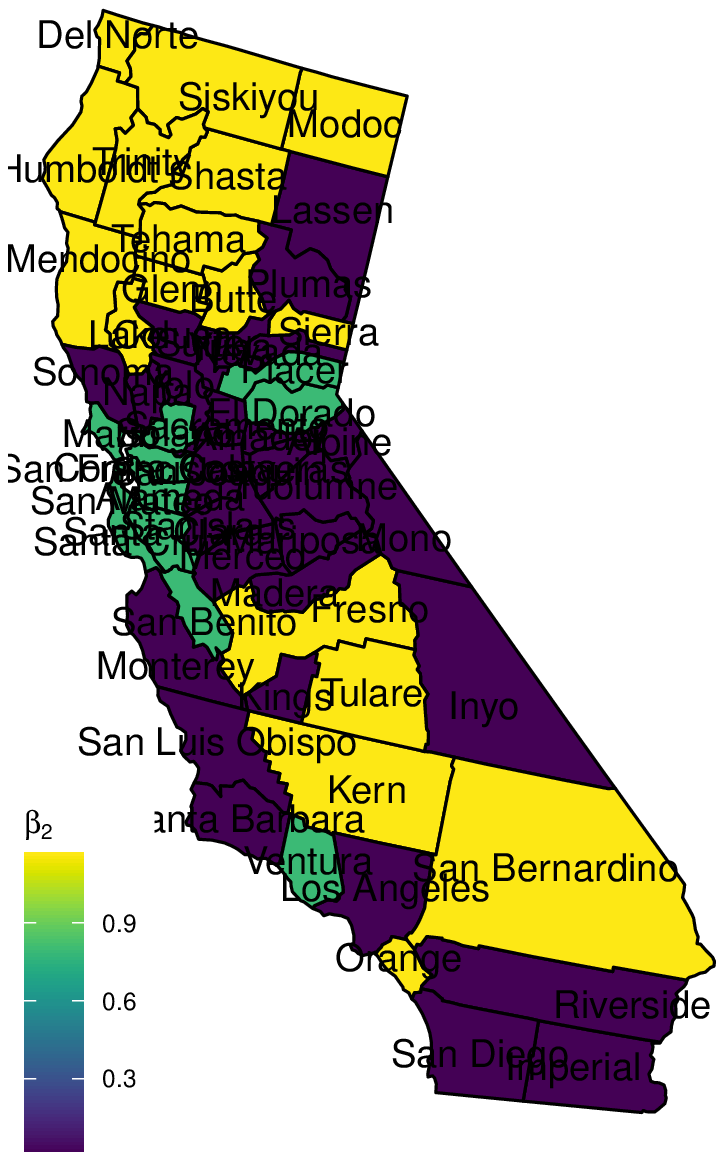}
\endminipage\hfill
\minipage{0.4\textwidth}
  \includegraphics[width=\linewidth]{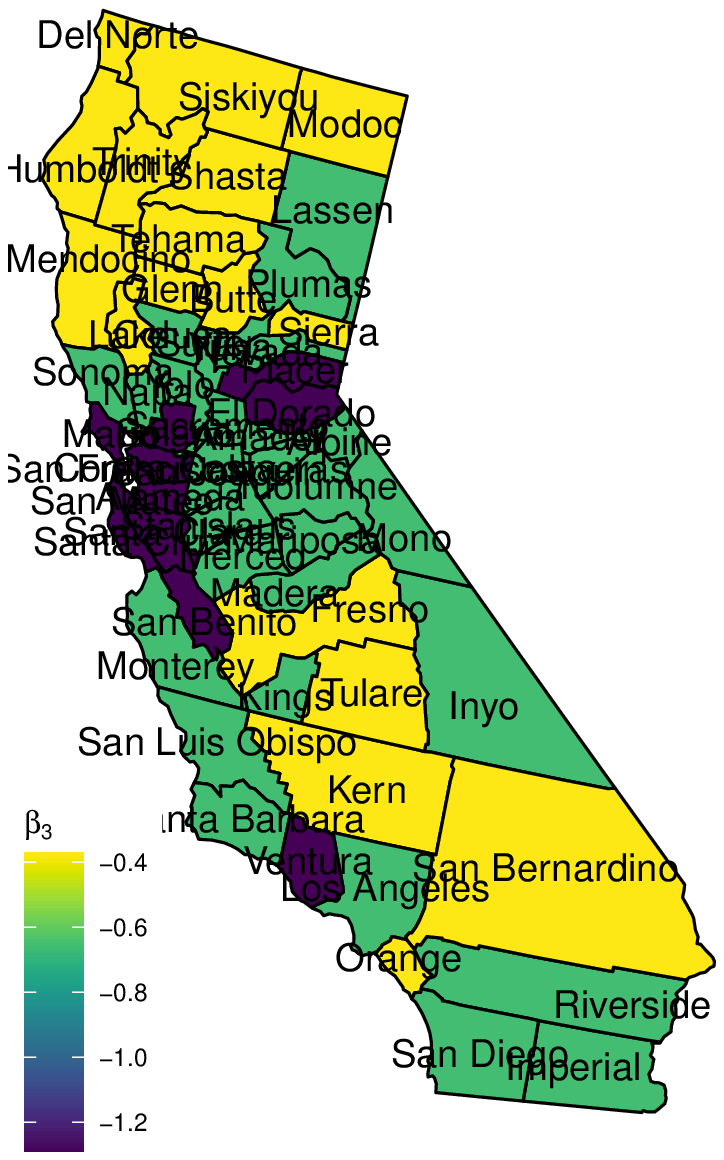}
\endminipage\hfill
\minipage{0.2\textwidth}
  \includegraphics[width=\linewidth]{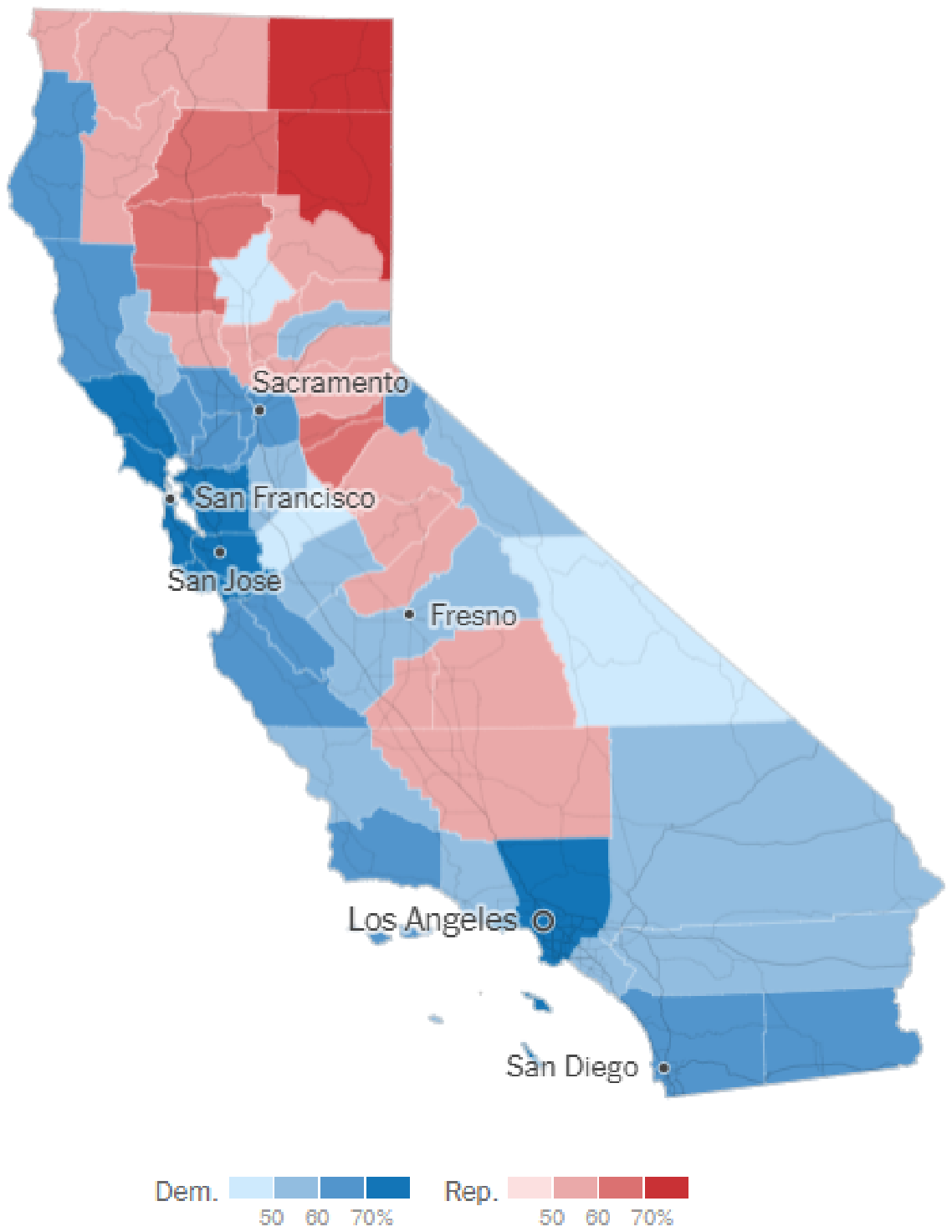}
\endminipage\hfill

\caption {\label{fig:4} Top left: The reporting partition; Top right: Visualized intercept; Bottom left: Visualized slope for GDP; Bottom middle: Visualized slope for log odds of the unemployment rate; Bottom right: 2020 presidential election results map (source: \url{https://www.nytimes.com/interactive/2020/11/03/us/elections/results-california.html})}
\end{figure}

We apply our proposed method under the same setting as described in Section \ref{sec 5.1: Precipitation}. The spatial smoothness parameter is selected as $\lambda=0.1$ based on the maximum marginal likelihood. The average rand index between the final cluster assignment and the ones from 100 replications is 0.9114, which confirms that the final cluster partition is representative. We present the clustering map in Figure \ref{fig:4} and summarize the regression parameter estimates for each of the three clusters in Table \ref{tab:3}. From the results, we can see a uniform pattern that the annual household median income is negatively associated with the unemployment rate and positively associated with the GDP in all three clusters, which agrees with the common sense. Among the obtained three clusters, Cluster 1 (see Figure \ref{fig:4}) has the strongest negative association between the unemployment and the median income; and it can be observed that most of the counties in the bay area (including Santa Clara and San Mateo) belong to this cluster. For Cluster 3, in which GDP has the lowest impact on the household income, most counties in this cluster are blue counties (Democrats votes $\geq 60\%$ during the 2020 presidential election), including Napa, Sonoma, Yolo, Santa Barbara, Los Angeles, San Diego, and Imperial. Those results suggest that political opinions and industrial structure may be potential confounders that can be included in the future analysis.

\begin{table}[htp]
    \centering
        \caption{\label{tab:3} Cluster-wise parameter estimates for the income data}
   \begin{tabular}{ccccccc}
    \toprule 
Cluster & $\hat{\beta}_{\text{intercept}}$ & $\hat{\beta}_{\text{GDP}}$ & $\hat{\beta}_{\text{unemployment}}$ & $\hat{\sigma^2}$ & $\hat{\ell}$ & $\hat{\alpha}$\\
    \midrule 
1& -1.761 & 0.804 & -1.291 & 0.246 & 1.789 & 0.082\\
2& -1.195 & 1.167 & -0.368 & 0.123 & 2.360 & 0.041\\
3& -1.454 & 0.016 & -0.650 & 0.147 & 8.284 & 0.113\\
    \bottomrule
    \end{tabular}
\end{table}

\section{Discussion}\label{sec 6: Discussion}
In this paper, we propose a general Bayesian spatial clustering method based on the product partition model equipped with a Markov random field structure for panel data analysis. We study the fundamental properties of MRF-PPM, and prove a weak clustering consistency result under mild conditions on the MRF structure. A computationally tractable MCMC algorithm, as well as a model selection method based on the marginal likelihood are introduced. Numerical studies  confirm that MRF-PPM effectively avoids the over-clustering issue and is more robust to model mis-specification compared to the classical PPM.  

Several future work directions remain open. First, it is difficult to study the asymptotic behavior of MRF-PPM prior when $N\to\infty$, as Kolmogorov's extension theorem does not hold generally after accounting for spatial information. It will be of interest to prove a Bayesian clustering consistency result when $N\to\infty$ as obtained in \citet{su2016identifying} and \citet{bonhomme2015grouped} for their frequentist approaches. Secondly, we assume that there is no temporal correlation between $Y_i(t_j^{(i)})$ and  $Y_i(t_k^{(i)})$ for $j \neq k$ when proving Theorem \ref{thm1} for general regression models. It will be of interest to relax this assumption. 
Our proposed prior can also be extended to allow a more generic form of the regression functions such as nonparametric or semi-parametric models. Developing efficient posterior computation and understanding the theoretical properties for this prior remains a changeling future work direction. 


\bibliographystyle{chicago}
\bibliography{Model.bib}

\end{document}